\documentclass[aoas,preprint]{imsart}

\usepackage{subfigure}
\usepackage{amsthm,amsmath,natbib}
\usepackage{graphicx}
\usepackage{bm,amssymb}
\usepackage{algpseudocode,algorithm}
\usepackage{color}
\RequirePackage[colorlinks,citecolor=blue,urlcolor=blue]{hyperref}
\usepackage{tikz}
\usetikzlibrary{patterns}
\usetikzlibrary{arrows}
\usepackage{multirow}
\usepackage{booktabs}

\arxiv{}

\setattribute{journal}{name}{}

\startlocaldefs

\newcommand{\ownint}[4]{{\int_{#1}^{#2} \! #3 \, \mathrm{d}#4}}
\newcommand{\tp}[1]{{#1^{\mathrm{T}}}}

\renewcommand{\P}{\operatorname{P\!}}
\newcommand{\diag}{\operatorname{diag}}

\newcommand{\pT}{p_\perp}

\endlocaldefs

\begin{document}

\begin{frontmatter}

\title{Shape-constrained uncertainty quantification in unfolding steeply falling elementary particle spectra}
\runtitle{Shape-constrained unfolding}


\author{\fnms{Mikael} \snm{Kuusela}\thanksref{t1,t2}\ead[label=e1]{mikael.kuusela@gmail.com}}
\address{Section de Math\'{e}matiques\\
\'{E}cole Polytechnique F\'{e}d\'{e}rale de Lausanne\\
EPFL Station 8, 1015 Lausanne\\
Switzerland\\
\printead{e1}}
\and
\author{\fnms{Philip B.} \snm{Stark}\ead[label=e2]{stark@stat.berkeley.edu}}
\address{Department of Statistics\\
University of California\\
Berkeley, CA 94720-3860\\
United States\\
\printead{e2}}
\affiliation{\'{E}cole Polytechnique F\'{e}d\'{e}rale de Lausanne and\linebreak University of California, Berkeley}

\thankstext{t1}{Supported by the Swiss National Science Foundation grant no.~200021-153595.}
\thankstext{t2}{Now at the University of Chicago.}

\runauthor{M. Kuusela and P.B. Stark}

\begin{abstract}
The high energy physics unfolding problem is an important statistical inverse problem in data 
analysis at the Large Hadron Collider (LHC) at CERN. 
The goal of unfolding is to make nonparametric inferences about a particle spectrum 
from measurements smeared by the finite resolution of the particle detectors. 
Previous unfolding methods use ad hoc discretization and regularization,
resulting in confidence intervals that can have significantly lower coverage than their nominal
level. 
Instead of regularizing using a roughness penalty or stopping iterative methods early, 
we impose physically motivated shape constraints: positivity, monotonicity, and convexity.
We quantify the uncertainty by constructing a nonparametric confidence 
set for the true spectrum, consisting of all those spectra that satisfy the shape 
constraints and that predict the observations within an appropriately calibrated level of fit. 
Projecting that set produces simultaneous confidence intervals for all functionals of the spectrum, including averages within bins. 
The confidence intervals have guaranteed conservative frequentist finite-sample coverage in the important and challenging class of unfolding problems for steeply falling particle spectra. 
We demonstrate the method using simulations that mimic unfolding
the inclusive jet transverse momentum spectrum at the~LHC.
The shape-constrained intervals provide usefully tight conservative inferences, 
while the conventional methods suffer from severe undercoverage.
\end{abstract}



\end{frontmatter}

\section{Introduction} \label{sec:intro}

This paper studies shape-constrained statistical inference in the high energy physics (HEP)
\emph{unfolding problem} \citep{Prosper2011,Cowan1998,Blobel2013,Kuusela2015}. 
This is an important statistical inverse problem arising 
at the Large Hadron Collider (LHC) at CERN, the European Organization for Nuclear Research,
near Geneva, Switzerland.
LHC measurements are affected by the finite resolution of the particle detectors. 
This causes the observations to be ``smeared'' or ``blurred'' versions of the true physical spectra. 
The unfolding problem is the inverse problem of making inferences about the true spectrum from the
noisy, smeared observations. 

The unfolding problem can be formalized using 
indirectly observed Poisson point processes \citep{Kuusela2015}.
It is an ill-posed Poisson inverse problem \citep{Antoniadis2006,Reiss1993}. 
Let $M$ and $N$ be Poisson point processes with intensity functions 
$f_0$ and $g_0$ and state spaces $T$ and $S$, respectively. 
We assume that the state spaces are compact intervals of real numbers.
Depending on the analysis, they could represent the energies, momenta, or production 
angles of particles, for example.
Let $M$ represent the true particle-level collision events, and $N$ the smeared detector-level events.
The intensity functions $f_0$ and $g_0$ are related by the Fredholm integral equation
\begin{equation}
   g_0(s) = (Kf_0)(s) = \ownint{T}{}{k(s,t)f_0(t)}{t}, \quad s \in S. \label{eq:intEq}
\end{equation}
The kernel $k(s,t)$, which represents the response of the particle detector, is
\begin{equation}
    k(s,t) \equiv p(Y=s|X=t, X \text{ observed}) P(X \text{ observed} | X = t), \label{eq:kernelDef}
\end{equation}
where $X$ is a physical particle-level event and $Y$ the corresponding smeared detector-level event. 
In this paper, we take $k(s,t)$ to be known. 
The goal of unfolding is to make inferences about the true intensity $f_0$ 
from a single realization of the smeared process $N$.

Unfolding is used in dozens of LHC analyses annually.
Existing unfolding methods, implemented 
in the {\sc RooUnfold} \citep{Adye2011PHYSTAT} software framework,
have serious practical limitations.  
These methods first discretize the continuous problem in 
Equation~\eqref{eq:intEq} using histograms 
\citep[Chapter~11]{Cowan1998} and then 
regularize the estimation problem either by stopping the expectation-maximization (EM)
iteration early \citep{DAgostini1995} or by using certain variants of Tikhonov regularization \citep{Hoecker1996,Schmitt2012}. 
Both approaches are ad hoc and produce biased estimates with formal uncertainties 
that can grossly underestimate the true uncertainty.

More specifically, let $\{T_i\}_{i=1}^p$ and $\{S_i\}_{i=1}^n$ be 
partitions of the true space~$T$ and 
the smeared space $S$ using histogram bins, and let
\begin{equation*}
 \bm{\lambda}_0 \equiv \tp{\bigg[\ownint{T_1}{}{f_0(t)}{t},\ldots,\ownint{T_p}{}{f_0(t)}{t}\bigg]} \text{ and }   
 \bm{\mu}_0 \equiv \tp{\bigg[\ownint{S_1}{}{g_0(s)}{s},\ldots,\ownint{S_n}{}{g_0(s)}{s}\bigg]}
\end{equation*}
be the expected number of true events and of smeared events in each bin, respectively. 
The two are related by $\bm{\mu}_0 = \bm{K}\bm{\lambda}_0$, 
where the elements of the response matrix $\bm{K}$ are
\begin{equation}
 K_{i,j} \equiv \frac{\ownint{S_i}{}{\ownint{T_j}{}{k(s,t)f_0(t)}{t}}{s}}{\ownint{T_j}{}{f_0(t)}{t}}, \quad i = 1,\ldots,n, \quad j = 1, \ldots, p. \label{eq:smearingMatrix}
\end{equation}
This matrix depends on the shape of $f_0$ within each true bin. 
Since $f_0$ is unknown, the matrix $\bm{K}$ is in practice constructed by replacing $f_0$ with a simulated ansatz obtained using a Monte Carlo (MC) event generator. After discretization, the unfolding problem amounts to estimating $\bm{\lambda}_0$ in the Poisson 
regression problem $\bm{y} \sim \nobreak \mathrm{Poisson}(\bm{K}\bm{\lambda}_0)$, where 
$\bm{y} \equiv \tp{\left[N(S_1), \ldots, N(S_n)\right]}$ is the 
vector of bin counts corresponding to
the histogram of smeared observations.

The Tikhonov-regularized unfolding techniques then make a Gaussian approximation to the 
Poisson likelihood and estimate $\bm{\lambda}_0$ by solving the optimization problem
\begin{equation}
   \min_{\bm{\lambda} \in \mathbb{R}^p} \tp{(\bm{y}-\bm{K}\bm{\lambda})} \bm{\hat{C}}^{-1} (\bm{y}-  
   \bm{K}\bm{\lambda}) + \delta P(\bm{\lambda}), \label{eq:Tikhonov}
\end{equation}
where $\bm{\hat{C}} \equiv \mathrm{diag}(\bm{y})$ is an estimate of the covariance of 
$\bm{y}$, $\delta > 0$ is a regularization parameter, and $P(\bm{\lambda})$ 
is a penalty term to regularize the otherwise ill-posed problem.
Two Tikhonov-regularized unfolding techniques are common in LHC data analysis: 
The {\sc TUnfold} variant \citep{Schmitt2012} uses the penalty term 
$P(\bm{\lambda}) = \|\bm{L}(\bm{\lambda}-\bm{\lambda}^\mathrm{MC})\|_2^2$, where 
$\bm{\lambda}^\mathrm{MC}$ is a Monte Carlo prediction of $\bm{\lambda}_0$ and 
$\bm{L}$ is typically a discretized second-derivative operator. 
The singular value decomposition (SVD) variant \citep{Hoecker1996} 
replaces the difference $\bm{\lambda}-\bm{\lambda}^\mathrm{MC}$ with the binwise ratios of 
$\bm{\lambda}$ and $\bm{\lambda}^\mathrm{MC}$. 
EM with early stopping \citep{DAgostini1995} starts the 
iteration from~$\bm{\lambda}^\mathrm{MC}$ and then stops before convergence, which regularizes 
the solution (iterating to convergence produces undesired oscillations). 
All these methods regularize the problem by biasing the solution towards a 
Monte Carlo prediction of~$\bm{\lambda}_0$. 
The uncertainty of the resulting point estimate $\bm{\hat{\lambda}}$ is 
then quantified by estimating its binwise standard errors using error propagation, 
ignoring the discretization and regularization biases.

The biggest problem with the current unfolding methods is that their uncertainty estimates are
unrealistically small.
In HEP applications, it is crucial to have confidence intervals with good frequentist coverage properties. 
But it is well understood that constructing confidence intervals using only the variability of a point estimate generally leads to undercoverage if the point estimate is biased.
These issues have been studied extensively, for instance in the spline smoothing literature 
\citep[Chapter 6]{Ruppert2003}.
\citet{Kuusela2015} show that standard 
uncertainty quantification techniques can lead to dramatic undercoverage in the unfolding problem. 
Moreover, using a Monte Carlo prediction first to discretize and then to regularize the problem 
introduces an unnecessary model-dependence and an additional uncertainty that is extremely difficult 
to quantify rigorously.

Spectra that fall steeply over many orders of magnitude are especially difficult to estimate well.
For such spectra, the response matrix $\bm{K}$ depends strongly
on the assumed Monte Carlo model, since the spectrum varies substantially within each true bin.
Current techniques regularize estimates of such spectra by requiring the unfolded solution 
to be close to a steeply falling Monte Carlo prediction. 
Trying to reduce the Monte Carlo dependence by using a global second-derivative 
penalty that does not depend on 
$\bm{\lambda}^\mathrm{MC}$ does not work well because the true solution has a highly
heterogeneous second derivative, large on the left and small on the right.

Energy, invariant mass and transverse momentum spectra of particle interactions are typically
steeply falling (assuming that the analysis is carried out above relevant low-energy thresholds). 
Recent LHC analyses that involve unfolding steeply falling spectra include the 
measurement of the differential cross section of jets \citep{CMS2013IncJets}, 
top quark pairs \citep{CMS2013TopPairs}, 
the $W$ boson \citep{ATLAS2012W}, and the Higgs boson \citep{CMS2016Higgs}. 
Expected key outcomes of LHC Run~II, currently underway, include more precise 
measurements of these and related steeply falling differential cross sections. 
A better method for unfolding such spectra is needed.

\begin{figure}[!t]
\includegraphics[trim = 0cm 0cm 0cm 0cm, clip=true, width=9cm]{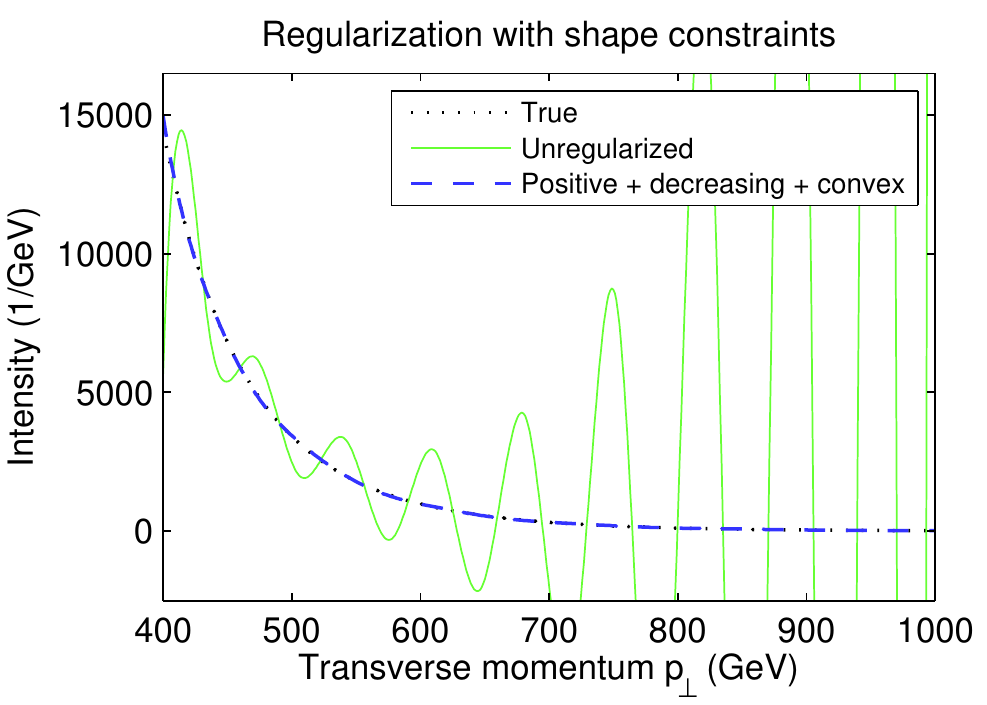}
\caption{Regularization of a steeply falling elementary particle spectrum using shape constraints. Here the spectrum is the inclusive jet differential cross section described in Section~\ref{sec:setup}. Unfolding these data is an ill-posed problem as demonstrated by the solid curve, which shows a typical unregularized point estimate (here a cubic spline that maximizes the likelihood). Shape constraints provide a natural way to rule out such unphysical solutions. If the spline coefficients are constrained so that the estimate is positive, decreasing and convex (dashed line), it becomes hard to distinguish from the true spectrum (dotted line). (These point estimates are presented to demonstrate how well shape constraints regularize the ill-posed problem. The uncertainty quantification method proposed in this paper
does not involve point estimates or splines.)}
\label{fig:introShapeConst}
\end{figure}

This paper develops a new method for forming rigorous confidence
intervals in the unfolding problem using an approach that is particularly well suited 
to steeply falling spectra. 
Instead of regularizing using a roughness penalty or by stopping EM early, we use 
shape constraints on the spectrum: positivity, monotonicity, and convexity.
Such constraints are an intuitive and physically justified way to impose regularity in a
wide range of unfolding analyses, including those mentioned above, 
without the need to bias the solution towards a Monte Carlo prediction nor to choose a 
roughness measure or regularization parameter. 
The use of shape constraints to estimate a steeply falling spectrum is 
demonstrated in Figure~\ref{fig:introShapeConst}.

\begin{figure}[!t]
	\includegraphics[trim = 0cm 0cm 0cm 0cm, clip=true, width=9cm]{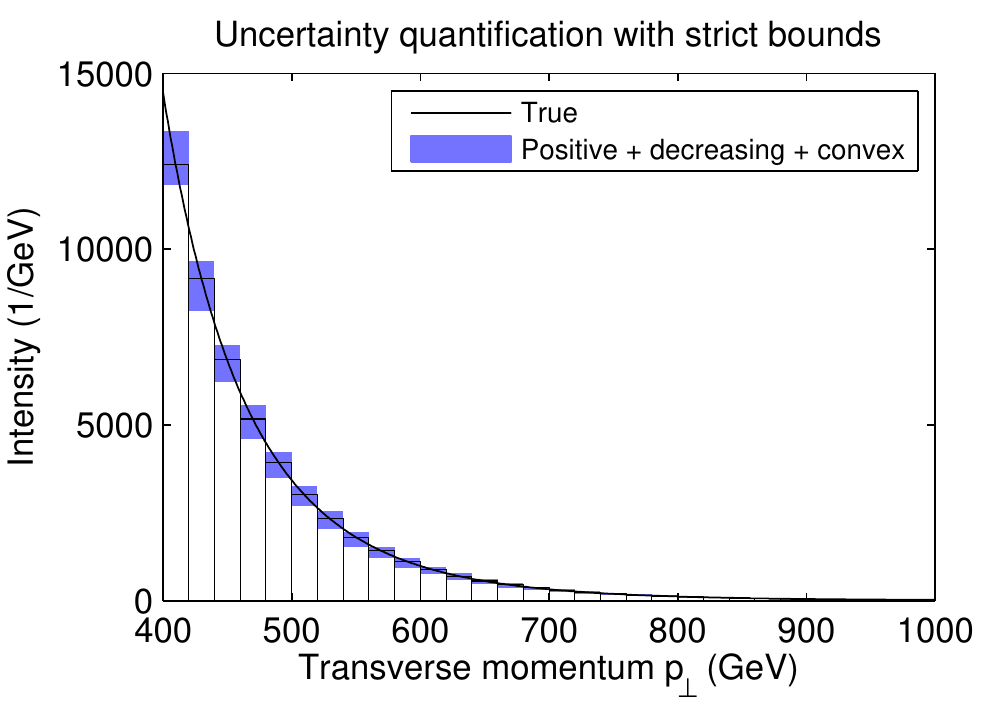}
	\caption{Shape-constrained strict bounds for unfolding the inclusive jet transverse momentum spectrum described in Section~\ref{sec:setup}. The confidence intervals are formed by considering all functions that satisfy the shape constraints (here positivity, monotonicity, and convexity) and that fit the observations within a tolerance calibrated to give the specified simultaneous confidence level (here 95\:\%).
        By construction, the finite-sample simultaneous coverage of these intervals is guaranteed to be at least 95\:\%.}
	\label{fig:introUQ}
\end{figure}

Shape constraints are easy to incorporate into uncertainty quantification 
using \emph{strict bounds} \citep{Stark1992}, which construct simultaneous confidence intervals by solving optimization problems
over the set of all functions that satisfy the shape constraints 
and agree adequately with the data (the misfit criterion and misfit level are chosen to yield the desired coverage probability). 
For the unfolding problem, the intervals are computed using a semi-discrete 
forward mapping, allowing the spectrum to vary arbitrarily within the true
bins, subject only to the shape constraints. 
This approach enables us to form confidence intervals for $\bm{\lambda}_0$ 
with \emph{guaranteed finite-sample simultaneous frequentist coverage},
avoiding the model-dependence and discretization and regularization biases
of existing methods. The proposed intervals are conservative, but still usefully sharp. Figure~\ref{fig:introUQ} shows an example of the resulting intervals for the inclusive jet transverse momentum 
spectrum studied in Section~\ref{sec:simulations}. 
Deriving such intervals and methods to compute them
are the main contributions of this paper. 
To do so, we extend the strict bounds methodology of \citet{Stark1992} to 
Poisson noise and develop a new way of imposing monotonicity 
and convexity constraints conservatively. 

The rest of this paper is structured as follows:
Section~\ref{sec:context} sketches the physical and statistical background of this work.
Section~\ref{sec:outline} provides an outline the proposed inference method.
Section~\ref{sec:SCU} explains in detail how to construct and compute the shape-constrained strict bounds. 
Section~\ref{sec:simulations} illustrates the bounds in a simulation study and
demonstrates that the existing methods can fail to achieve their nominal coverage,
even in a typical unfolding scenario. 
Section~\ref{sec:discussion} concludes.
Appendices give derivations and implementation~details.

\section{Context and background} \label{sec:context}

\subsection{LHC experiments and unfolding}

The Large Hadron Collider (LHC)
is the world's largest and most powerful particle accelerator. 
It collides protons at velocities close to the speed of light
to study the properties and interactions of elementary particles created in these collisions. 
The collision events are recorded using four massive underground particle detectors, called 
ATLAS, ALICE, CMS, and LHCb. 
ATLAS and CMS are general-purpose experiments, while ALICE focuses on heavy-ion and 
LHCb on bottom-quark physics. 
Together, these experiments produce approximately 30~petabytes of data annually.

\begin{figure}[t]
\includegraphics[trim = 0cm 0cm 0cm 0cm, clip=true, width=12.5cm]{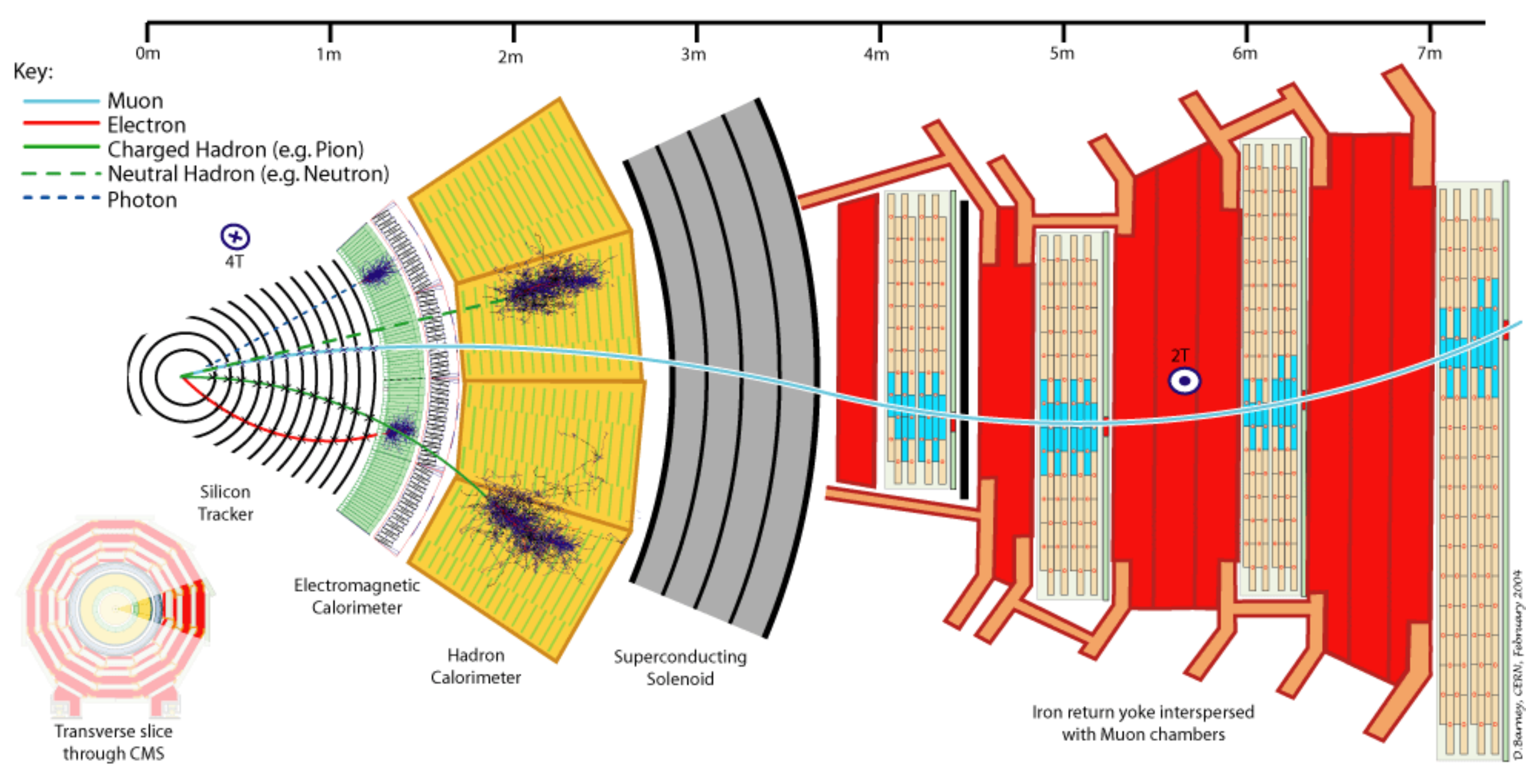}
\caption[]{Structure of the CMS experiment at the LHC~\citep{Barney2004}. 
The experiment uses layers of subdetectors to
identify the particles created in the proton-proton collisions and to estimate their properties. 
Copyright: CERN, for the benefit of the CMS Collaboration.}
\label{fig:CMS_slice}
\end{figure}

Figure~\ref{fig:CMS_slice} shows the structure of a typical HEP experiment 
(here the CMS detector, \citet{CMS2008JINST}). 
Particles collide at the center of the experiment. 
The collision point is surrounded by layers of detectors, 
which record information that can be used to reconstruct the collision event. 
The innermost layer is the \emph{tracker}, which measures the trajectories of charged particles. 
The next layer is the \emph{electromagnetic calorimeter}, 
which measures the energies of electrons, positrons, and photons. 
This is followed by the \emph{hadron calorimeter}, which 
measures the energies of hadrons---particles composed of quarks. 
The outermost layer consists of \emph{muon detectors}, which identify and trace muons. 
The whole detector system is in a strong magnetic field that bends the trajectories of charged particles, 
enabling their momenta to be measured. 
Each type of particle leaves a characteristic signature in the various subdetectors; 
see Figure~\ref{fig:CMS_slice}. 
Pattern recognition algorithms are then used to combine information from the subdetectors 
to reconstruct the collision event and to identify physics objects of interest.

The unfolding problem arises because of the finite resolution of the LHC particle detectors. 
Let $X_i$ be the true value of some quantity measured in the detector, for instance,
the energy of a particle hitting the electromagnetic calorimeter. 
Due to detector effects, the observed value of this quantity, say, $Y_i$, includes stochastic noise:
the conditional density of $Y_i$ given $X_i$, $p(Y_i|X_i)$, has support on a continuum of values 
around $X_i$. 
If the noise were additive, i.e., $Y_i = X_i + \varepsilon_i$, with $\varepsilon_i$ 
independent of $X_i$, then unfolding becomes a deconvolution problem. 
Usually the appropriate noise models are more complicated though,
and methods that can handle general smearing kernels are needed.

It follows from the law of rare events that the number of collisions 
in HEP experiments is, to a good approximation, Poisson distributed. 
The particle-level quantities $X_1, \ldots, X_\tau$ can be treated as a Poisson point process,
from which it follows that the smeared observations $Y_1,\ldots,Y_\xi$ also comprise 
a Poisson point process (we may have $\xi < \tau$ due to the limited efficiency of the detector). 
This leads to the indirectly observed Poisson point process model in
Section~\ref{sec:intro}; see also \citet[Section~3]{Kuusela2015}.

Unfolding is used extensively in LHC data analysis, but not all analyses need to be performed in the unfolded space.
In particular, searches for new particles and phenomena usually can be carried out as hypothesis tests
in the smeared space.
This is in contrast with \emph{measurement analyses} where the goal is to estimate some physically relevant
particle spectrum, which is usually most naturally done in the unfolded space. 
Typically, the main goal in
unfolding analyses is to estimate the spectrum in a form that allows other scientists to use it as
input to further analyses. This could include hypothesis tests in the unfolded space, combinations of unfolded
measurements, or inferring some physical parameter from one or more unfolded results.
For example, parton distribution functions \citep{Forte2013}, which 
quantify the internal structure of a proton, are estimated using simultaneous 
fits to several unfolded spectra \citep{NNPDF2015}.

A more detailed introduction to LHC experiments and the role of unfolding in their data analysis is given in 
\citet[Section~2]{Kuusela2015}.

\subsection{Previous work}

Shape-constrained statistical inference goes back to the seminal work of \citet{Grenander1956}, 
who derived the nonparametric maximum likelihood estimator of a density 
subject to a monotonicity constraint. 
There is now a large, diverse literature on the topic; see 
\citet{Robertson1988} and \cite{Groeneboom2014} for overviews. 
Interest has focused on shape-constrained point estimation of density or regression functions 
using direct observations, and on the nonstandard asymptotic properties of such estimators; see,
e.g., \citet{Groeneboom2001} and the references therein.
Similarly, there is an extensive literature on nonparametric deconvolution problems; see \citet{Meister2009} for a comprehensive introduction. 
But using shape constraints to regularize deconvolution-type problems has received only
limited attention; examples include 
\citet{Wahba1982,Stark1992,Carroll2011,Pflug2013}.

Constructing nonparametric confidence intervals for functions is challenging. 
When the intervals are based on smoothed estimators, it is difficult to take into account the 
regularization bias~\citep[Chapter 6]{Hall2013,Ruppert2003}. 
In fact, under the usual assumptions it is not possible to obtain nonparametric correct-coverage confidence bands that adapt to the smoothness of the underlying function \citep{Low1997,Genovese2008}. This difficulty can be circumvented using shape constraints
\citep{Low1997,Cai2013}, but so far most 
approaches have been asymptotic \citep{Banerjee2001,Groeneboom2015}. 
For direct observations, finite-sample shape-constrained confidence intervals with various degrees of adaptivity have been obtained by \citet{Hengartner1995,Dumbgen1998,Dumbgen2003,Davies2009,Cai2013}. 
\citet{Rust1972} sketch a construction similar to ours for indirect observations, but do not work out the details or provide a concrete application. In follow-up work, \citet{Pierce1985} provide an algorithm for various shape constraints, but only for a discretized version of the nonparametric problem. \citet{Stark1992} provides a general recipe for constructing constrained finite-sample confidence intervals in indirect nonparametric problems and serves as the methodological basis for our work.

The two main approaches to unfolding in experimental HEP are Tikhonov regularization \citep{Hoecker1996,Schmitt2012}, which goes back to the work of \citet{Tikhonov1963} and \citet{Phillips1962},
 and D'Agostini iteration \citep{DAgostini1995}, which applies EM \citep{Dempster1977} in the discrete Poisson regression model but stops before convergence, to provide regularization. 
D'Agostini iteration has been discovered before in various fields, including optics \citep{Richardson1972}, astronomy \citep{Lucy1974}, and positron emission tomography \citep{Shepp1982,Lange1984,Vardi1985}. 
It was popularized in HEP by \citet{DAgostini1995}, but had already been studied earlier by \citet{Kondor1983} and \citet{Muelthei1987MMAS, Muelthei1987NIM, Muelthei1989}. Recent proposals for solving the unfolding problem include Bayesian estimation \citep{Choudalakis2012}, empirical Bayes estimation \citep{Kuusela2015}, and EM iteration with smoothing \citep{Volobouev2015}. 
The last two papers also study the coverage properties of confidence intervals for the unfolded spectrum. (\citet{DAgostini1995} also describes a smoothed version of the EM iteration, but the method is typically applied without smoothing in current LHC analyses.)

Strict bounds confidence intervals have been used primarily in geophysics, solar physics, and density estimation,
using constraints of positivity, monotonicity, $m$-modality, and general conical constraints; see \citet{Stark1992,Hengartner1992,Hengartner1995} and the references therein. A similar approach was developed by \citet{Burrus1965} for applications in nuclear spectroscopy. 
\citet{Burrus1965,Burrus1969,OLeary1986,Rust1994} use an approach related to ours for constructing positivity-constrained unfolded confidence intervals. 
Their applications use only a positivity constraint because the underlying the spectra contain one or more peaks.
In contrast, for differential cross section measurements in HEP, the spectra typically are known to be 
steeply falling and hence monotone and usually also convex. 
Earlier work on strict bounds also does not explicitly treat Poisson noise.

\section{Outline of the methodology} \label{sec:outline}

We propose quantifying the uncertainty of the unfolded particle spectrum using shape-constrained \emph{strict bounds confidence intervals} \citep{Stark1992}. A rough outline of the approach is as follows:
\begin{enumerate}
	\item For fixed $\alpha \in (0,1)$, form a $1-\alpha$ simultaneous confidence set in the binned smeared space. This is the set of all those smeared histograms that are compatible with the smeared observations.
	\item Consider the preimage of this set in the unbinned true space. The preimage is a $1-\alpha$ simultaneous confidence set for the true spectrum. Intuitively, it is the set of all those unfolded spectra that are consistent with the smeared observations.
	\item Take the intersection of this unfolded confidence set with the set of spectra that satisfy the shape constraints. If the true spectrum satisfies the imposed constraints, the intersection remains a $1-\alpha$ simultaneous confidence set for that spectrum, because spectra that violate the constraints cannot contribute to the coverage. The constraints we consider are positivity; positivity and monotonicity; and positivity, monotonicity, and convexity.
	\item Characterize the unbinned shape-constrained confidence set in terms of its projections to given functionals. For any functional, the interval from the smallest value of the functional over the confidence set to the largest value of the functional over the confidence set is a $1-\alpha$ confidence interval for that functional.  Moreover, for any set of functionals whatsoever (even an uncountably infinite set), those intervals are simultaneous $1-\alpha$ confidence intervals for the functionals, because all the intervals cover whenever the unbinned confidence set covers the true unfolded spectrum---which occurs with probability at least $1-\alpha$. In particular, one may take the functionals to be a collection of binwise integrals of the spectrum, which yields confidence intervals for binwise averages, with $1-\alpha$ simultaneous coverage in the binned true space.
\end{enumerate}
The last step of this procedure is conservative and may lead to overcoverage. However, undercoverage, which could lead to seriously faulty inferences, is not possible. The rest of this section develops this approach more rigorously, specifically for Poisson unfolding.

Let $F$ denote the space of (sufficiently regular) functions on the compact interval 
$T \subset \mathbb{R}$. 
The random point measure $M$ on state space $T$ is a \emph{Poisson point process} 
\citep{Reiss1993} \emph{with a nonnegative intensity function} $f \in F$ if and only if
\begin{itemize}
\item[(a)] $M(B) \sim \mathrm{Poisson}(\lambda(B))$ with $\lambda(B) = \ownint{B}{}{f(t)}{t}$ 
for every Borel set $B \subset T$;
\item[(b)] $M(B_1), \ldots, M(B_n)$ are independent for pairwise disjoint 
Borel sets $B_i \subset T, \, i=1, \ldots, n$.
\end{itemize}
In the unfolding problem, the true particle-level events are a realization of such a Poisson point
process $M$ on state space $T$ with intensity function $f_0$.
Let $G$ denote the space of (sufficiently regular) functions on the compact interval 
$S \subset \mathbb{R}$.  
The smeared detector-level events are a realization of the 
Poisson point process $N$ on state space $S$ with a nonnegative intensity function $g_0 \in G$. 
The two intensity functions are related by $g_0 =\nobreak Kf_0$, where the forward operator $K:F \rightarrow G$ is defined by Equation~\eqref{eq:intEq}.

HEP data are typically binned for convenience and computational tractability.
For steeply falling spectra, the leftmost bins may contain billions of observations:
treating them individually is not currently computationally feasible. 
We assume therefore that the smeared observations are counts in bins 
$\bm{y} \equiv \tp{\left[N(S_1),\ldots,N(S_n)\right]}$, where $\{S_j\}_{j=1}^n$ is a 
binning of the smeared space $S$:
\begin{equation}
 S_j \equiv \begin{cases} [S_{j,\mathrm{min}},S_{j,\mathrm{max}}), & j=1,\ldots,n-1, \\ [S_{j,\mathrm{min}},S_{j,\mathrm{max}}], & j=n, \end{cases}
\end{equation}
with $S_{j,\mathrm{max}} = S_{j+1,\mathrm{min}},\,j=1,\ldots,n-1$. 
The expected value of these counts is 
$\bm{\mu}_0 \equiv \tp{\left[\ownint{S_1}{}{g_0(s)}{s},\ldots,\ownint{S_n}{}{g_0(s)}{s}\right]}$, 
which has components
\begin{equation}
 \mu_{0,j} \equiv \ownint{S_j}{}{\ownint{T}{}{k(s,t)f_0(t)}{t}}{s} = \ownint{T}{}{k_j(t)f_0(t)}{t} \equiv K_j f_0, \quad j=1,\ldots,n,
\end{equation}
where $k_j(t) \equiv \ownint{S_j}{}{k(s,t)}{s}$. 
We assume here and below that $f_0$ and $k$ are sufficiently regular that we can change the
order of integration.

Let $\mathcal{K}$ denote the semi-discrete linear operator that maps a particle-level intensity function
$f$ to the detector-level mean vector $\bm{\mu}$:
\begin{equation}
 \mathcal{K}: F \rightarrow \mathbb{R}^n,\ f \mapsto \tp{\left[K_1 f,\ldots,K_n f\right]}.
\end{equation}
With this notation, our statistical model is
\begin{equation}
 \bm{y} \sim \mathrm{Poisson}(\bm{\mu}_0), \text{ with } \bm{\mu}_0 = \mathcal{K}f_0,
\end{equation}
with the components of $\bm{y}$ independent.

We seek simultaneous confidence intervals for the binned expected value vector of the true process~$M$:
\begin{equation}
\bm{\lambda}_0 \equiv \tp{\left[\ownint{T_1}{}{f_0(t)}{t},\ldots,\ownint{T_p}{}{f_0(t)}{t}\right]} = \tp{\left[H_1 f_0,\ldots,H_p f_0\right]}, \label{eq:lambda}
\end{equation}
where $\{T_k\}_{k=1}^p$ is a partition of $T$ into intervals and, for each $k=1,\ldots, p$,  
$H_kf \equiv \ownint{T_k}{}{f(t)}{t}$. 
As before, the partition is of the form
\begin{equation}
 T_k \equiv \begin{cases} [T_{k,\mathrm{min}},T_{k,\mathrm{max}}), &k=1,\ldots, p-1, \\ [T_{k,\mathrm{min}},T_{k,\mathrm{max}}], & k=p, \end{cases}
\end{equation}
with $T_{k,\mathrm{max}} = T_{k+1,\mathrm{min}},\, k=1, \ldots, p-1$.

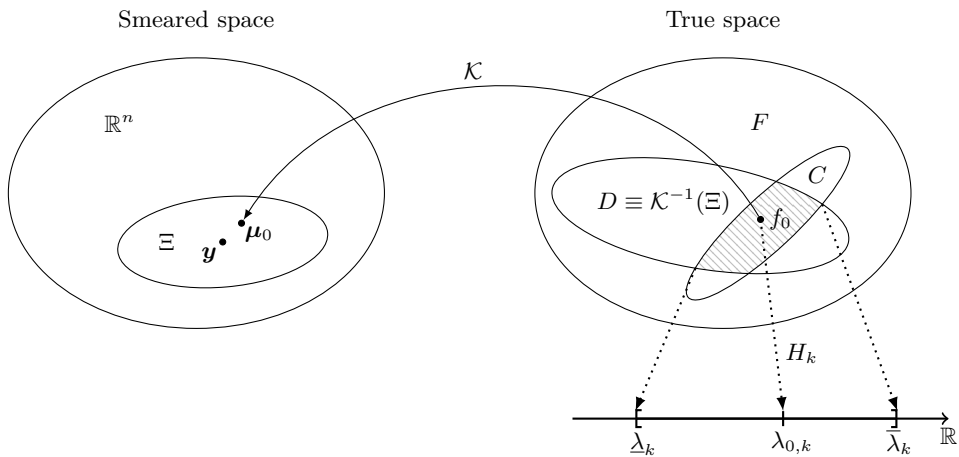
\begin{figure}[!t]

\begin{center}
\begin{tikzpicture}
\draw (0,0) circle[x radius = 2.5, y radius = 1.8] node[above left, outer sep = 20]{$\mathbb{R}^n$} node[above, outer sep = 58]{Smeared space};
\draw (7,0) circle[x radius = 2.5, y radius = 1.8] node[above, outer sep = 21]{$\quad \qquad F$} node[above, outer sep = 58]{True space};

\begin{scope}
\clip (6.7,-0.3) circle[x radius = 2, y radius = 0.7, rotate = -10];
\clip (7.6,-0.4) circle[x radius = 1.45, y radius = 0.35, rotate = 43];
\draw[pattern = north west lines, pattern color = lightgray] (6,-2) rectangle (9,2);
\end{scope}

\draw (0.35,-0.65) circle[x radius = 1.4, y radius = 0.6, rotate = 5] node[circle,draw,fill=black,inner sep = 0.8]{} node[below left, inner sep = 2]{$\bm{y}$} node[left, outer sep = 15]{$\Xi$};
\draw (6.7,-0.3) circle[x radius = 2, y radius = 0.7, rotate = -10] node[above, inner sep = 0]{$D \equiv \mathcal{K}^{-1}(\Xi) \qquad \quad$};
\draw (7.6,-0.4) circle[x radius = 1.45, y radius = 0.35, rotate = 43] node[above right,outer sep = 12]{$C$};

\node[circle,draw,fill=black,inner sep = 0.8](f) at (7.5,-0.35){};
\node[circle,draw,fill=black,inner sep = 0.8](mu) at (0.6,-0.4){};
\draw[-latex] (f) node[right, inner sep = 3]{$f_0$} ..controls (6.1,2) and (2.1,2) .. (mu) node[pos = 0.55,above]{$\mathcal{K}$} node[below right, inner sep = 1]{${\bm{\mu}}_0$};

\draw[->, thick] (4+1,-2.5-0.5) -- (10,-2.5-0.5) node[below]{$\mathbb{R}$};
\draw[-latex,dotted, thick] (6.635,-1) -- (6.635-0.8,-2.4-0.5);
\draw[-latex,dotted, thick] (8.32,-0.15) -- (8.32+1,-2.4-0.5);
\draw[-latex,dotted, thick] (7.5,-0.35) -- (7.5+0.3,-2.4-0.5) node[pos = 0.7,right]{$H_k$};
\draw[{[-]},thick] (6.635-0.8,-2.5-0.5) node[below,outer sep = 2]{$\,\;\: \underline{\lambda}_k$} -- (8.32+1,-2.5-0.5) node[below,outer sep = 1]{$\,\overline{\lambda}_k$};
\draw[thick] (7.5+0.3,-2.4-0.5) -- (7.5+0.3,-2.6-0.5) node[below,inner sep = 1.7]{$\;\;\,\lambda_{0,k}$};

\end{tikzpicture}
\end{center}
    \vspace{-0.3cm}
    \caption{Sketch of strict bounds confidence intervals. 
    The set $\Xi$ in the smeared space is a confidence set for $\bm{\mu}_0$ based on $\bm{y}$.
    Its preimage $D \equiv \mathcal{K}^{-1}(\Xi)$ is a confidence set for $f_0$. 
    If $f_0$ satisfies the shape constraints $C$, then $C \cap D$ is also a confidence set for $f_0$.
    The endpoints of the confidence intervals are given by the extremal values of the functionals of interest $H_k$ over $C \cap D$.}
    \label{fig:strictBoundsConstruction}
\end{figure}

Instead of further discretizing the forward mapping,
we construct strict bounds confidence intervals \citep{Stark1992} under the semi-discrete mapping~$\mathcal{K}$
as follows (cf. the outline in the beginning of this section):
\begin{enumerate}
 
 \item For fixed $\alpha \in (0,1)$, let $\Xi \subset \mathbb{R}^n$ be a $1-\alpha$ simultaneous 
 confidence set for $\bm{\mu}_0$ in the smeared space, so that
 $\P_{f_0} \{ \Xi \ni \bm{\mu}_0 \}\ge 1-\alpha$.
  \item The preimage $D \equiv \mathcal{K}^{-1}(\Xi)$ of $\Xi$ under the mapping $\mathcal{K}$
 is a $1-\alpha$ confidence set for the unbinned particle-level intensity 
 function~$f_0$: $\P_{f_0} \{D \ni f_0\} \ge 1-\alpha$.
 (The symbol $D$ is mnemonic for \emph{data}.)
 
 \item Let $C \subset F$ denote the set of functions that satisfy the shape constraints.
 (The symbol $C$ is mnemonic for \emph{constraint}, and also for \emph{convex cone}, which
 all our constraint sets are.)
If $f_0 \in C$, then $C \cap D$ is also a $1-\alpha$ confidence set for $f_0$:
 $\P_{f_0} \{ C \cap D \ni f_0 \} \ge 1-\alpha$.
 
 \item Whenever $f_0 \in C \cap D$ (which occurs with probability at least $1-\nobreak\alpha$),
 any condition that holds for every $f \in C \cap D$ 
 holds for $f_0$. 
 In particular, $H_k f_0 \ge \inf_{f \in C \cap D} H_k f \equiv \underline{\lambda}_k$ 
 and $H_k f_0 \le \sup_{f \in C \cap D} H_k f \equiv \overline{\lambda}_k$,
 for every $k$.
 Each interval $[\underline{\lambda}_k,\overline{\lambda}_k]$ is a $1-\alpha$ confidence interval for~$\lambda_{0,k}$. 
 Moreover, the set $[\underline{\lambda}_1,\overline{\lambda}_1] \times \cdots \times [\underline{\lambda}_p,\overline{\lambda}_p] \subset \mathbb{R}^p$ is a $1-\alpha$ 
 simultaneous confidence set for $\bm{\lambda}_0$:
 Whenever $C \cap D$ covers $f_0$, the $p$ intervals 
 $\{ [\underline{\lambda}_j,\overline{\lambda}_j]\}_{j=1}^p$ all cover the corresponding components of 
 $\bm{\lambda}_0$ and $\P_{f_0}([\underline{\lambda}_1,\overline{\lambda}_1] \times \cdots \times [\underline{\lambda}_p,\overline{\lambda}_p] \ni \bm{\lambda}_0) \geq 1-\alpha$. 
\end{enumerate}
This construction is illustrated in Figure~\ref{fig:strictBoundsConstruction}.

The simultaneous coverage probability of $[\underline{\lambda}_1,\overline{\lambda}_1] \times \cdots \times [\underline{\lambda}_p,\overline{\lambda}_p]$ can be greater than that of $C \cap D$: it has elements that do not correspond to any $f \in \nobreak C \cap D$. Thus, the intervals are guaranteed to have simultaneous confidence level at least $1-\alpha$, but the actual coverage probability might be much~larger.

Constructing confidence intervals involves solving the optimization problems 
$\inf_{f \in C \cap D} H_k f$ and $\sup_{f \in C \cap D} H_k f$. 
Since
\begin{equation}
   \sup_{f \in C \cap D} H_k f = -\inf_{f \in C \cap D} -H_k f,
\end{equation}
we focus on the minimization problem, without loss of generality.
We cannot directly solve the infinite-dimensional optimization problem 
$\inf_{f \in C \cap D} H_k f$. 
Instead, we follow the approach of \citet{Stark1992} and use Fenchel
duality \citep[Section 7.12]{Luenberger1969} to bound $\inf_{f \in C \cap D} H_k f$ 
from below by a semi-infinite program with an $n$-dimensional unknown and an infinite set of 
constraints. 
We then discretize the constraints in such a way that every feasible point
of the discretized finite-dimensional dual program gives a lower bound for the value of the
original program $\inf_{f \in C \cap D} H_k f$,
ensuring that the simultaneous confidence level of the resulting intervals remains at least $1-\alpha$.

\section{Shape-constrained unfolding} \label{sec:SCU}

This section shows how to compute shape-constrained strict bounds confidence
intervals in the unfolding problem. 
Section~\ref{sec:smearedCS} develops the confidence set $\Xi$ for the smeared mean vector 
$\bm{\mu}_0$. 
Section~\ref{sec:duality} applies Fenchel duality to turn the infinite-dimensional program 
$\inf_{f \in C \cap D} H_k f$ into a semi-infinite program, whose explicit form for various shape 
constraints $C$ is derived in Section~\ref{sec:constraints}. 
Finally, Section~\ref{sec:discretization} explains how to discretize the 
semi-infinite program in such a way that the confidence level is preserved.

\subsection{Confidence set in the smeared space} \label{sec:smearedCS}

Constructing the confidence set $\Xi$ for the mean $\bm{\mu}_0$ in the model $\bm{y} \sim \mathrm{Poisson}(\bm{\mu}_0)$ is straightforward.
For fixed $\alpha \in (0,  1)$, set  $\alpha' = 1-(1-\alpha)^{1/n}$.
For each $j=1, \ldots, n$, 
let $[\underline{\mu}_{j,\alpha'},\overline{\mu}_{j,\alpha'}]$ be a 
$1-\alpha'$ confidence interval for $\mu_{0,j}$ constructed using $y_j$ only. 
Then 
\begin{align}
&\P_{f_0}(\underline{\mu}_{1,\alpha'} \leq \mu_{0,1} \leq \overline{\mu}_{1,\alpha'},\ldots,\underline{\mu}_{n,\alpha'} \leq \mu_{0,n} \leq \overline{\mu}_{n,\alpha'}) \\
&=  \prod_{j=1}^n \P_{f_0}(\underline{\mu}_{j,\alpha'} \leq \mu_{0,j} \leq \overline{\mu}_{j,\alpha'}) \geq  (1-\alpha')^n = 1 - \alpha, \notag
\end{align}
because the counts $y_j$ in disjoint
bins are independent.
Thus
\begin{equation}
\Xi \equiv [\underline{\mu}_{1,\alpha'},\overline{\mu}_{1,\alpha'}] \times \cdots \times [\underline{\mu}_{n,\alpha'},\overline{\mu}_{n,\alpha'}]
\end{equation}
is a $1-\alpha$ 
simultaneous confidence set for $\bm{\mu}_0$.

The binwise confidence intervals $[\underline{\mu}_{j,\alpha'},\overline{\mu}_{j,\alpha'}]$ can be formed using the Garwood construction \citep{Garwood1936}. 
For bin $j$, the interval endpoints are
\begin{eqnarray}
 \underline{\mu}_{j,\alpha'} &=&  \left \{ 
     \begin{array}{ll}  \frac{1}{2} F^{-1}_{\chi^2}\left(\frac{\alpha'}{2}; 2 y_j \right), & y_j > 0, \cr 0, & y_j = 0,
      \end{array} \right . \nonumber \\
       & \text{and} & \label{eq:GarwoodIntervals} \\
       \overline{\mu}_{j,\alpha'} & = &  \frac{1}{2} F^{-1}_{\chi^2}\left(1-\frac{\alpha'}{2}; 2(y_j+1) \right), \nonumber
\end{eqnarray}
where $F^{-1}_{\chi^2}(\,\cdot\,; k)$ is the inverse cumulative distribution function of the 
$\chi^2$ distribution with $k$ degrees of freedom. 
These intervals are guaranteed to have confidence level $1-\alpha'$, although the actual coverage probability is strictly greater than $1-\alpha'$ for any finite $\mu_{0,j}$, a consequence of the 
discreteness of the Poisson distribution.

It will be convenient to write the hyperrectangle $\Xi$ as a centered set: 
$\Xi = \{\bm{\tilde{y}} + \bm{\xi} \in \mathbb{R}^n: \|\diag(\bm{\ell})^{-1} \bm{\xi} \|_\infty \leq 1\}$, where, for each $j=1,\ldots,n$, 
$\tilde{y}_j \equiv (\underline{\mu}_{j,\alpha'}+\overline{\mu}_{j,\alpha'})/2$ and $\ell_j \equiv (\overline{\mu}_{j,\alpha'}-\underline{\mu}_{j,\alpha'})/2$.

\subsection{Strict bounds via duality} \label{sec:duality}

To compute the lower bound $v(\mathcal{P}) \equiv \inf_{f \in C \cap D} H_k f$, we follow the approach of \citet{Stark1992} and apply Fenchel duality. 
The Fenchel dual \citep[Section 7.12]{Luenberger1969} of the problem~is 
\begin{equation}
 v(\mathcal{D}) \equiv \sup_{f^* \in C^* \cap D^*} \left\{ \inf_{f \in D} f^*[f] + \inf_{f \in C} (H_k-f^*)[f] \right\},
\end{equation}
where 
\begin{align}
C^* &\equiv \left\{ f^* \in F^* : \inf_{f \in C} (H_k-f^*)[f] > -\infty \right\}, \\
D^* &\equiv \left\{ f^* \in F^* : \inf_{f \in D} f^*[f] > -\infty \right\},
\end{align}
and $F^*$ is the algebraic dual space of $F$, that is, the set of all linear functionals on $F$. 
The two problems satisfy weak duality, $v(\mathcal{P}) \geq v(\mathcal{D})$, and, when strict inequality holds, solving the dual problem gives a conservative lower bound. 
Under regularity conditions 
(see \citet[Section 7.12]{Luenberger1969}),
one can establish strong duality, $v(\mathcal{P}) = v(\mathcal{D})$, in which case the bound from the dual is tight.
(It is enough that $F$ is normed, the forward functionals 
$\{K_j\}$ are continuous, and the primal problem
is feasible; \linebreak see \citet[Section 10.1]{Stark1992}.)

The Fenchel dual can be written using a finite-dimensional unknown, which simplifies numerical solution:
the set $D^*$ consists of functionals that are linear combinations of the forward functionals 
$K_j$, 
\begin{equation}
D^* = \left\{ f^* \in F^* : f^* = \bm{\nu} \cdot \mathcal{K}, \;\bm{\nu} \in \mathbb{R}^n \right\},
\end{equation}
where 
$\bm{\nu} \cdot \mathcal{K} \equiv \sum_{j=1}^n \nu_j K_j$
\citep{Stark1992, Backus1970}.
The dual problem is therefore
\begin{equation}
 v(\mathcal{D}) = \sup_{\bm{\nu} \in \mathbb{R}^n : \bm{\nu} \cdot \mathcal{K} \in C^*} \left\{ \inf_{f \in D}  (\bm{\nu} \cdot \mathcal{K})[f] + \inf_{f \in C} (H_k - \bm{\nu} \cdot \mathcal{K})[f] \right\}.
\end{equation}
The first term can be expressed in closed form: when $f \in D$,
\begin{align}
 (\bm{\nu} \cdot \mathcal{K})[f] &= \tp{\bm{\nu}}(\bm{\tilde{y}} + \bm{\xi}) \geq \tp{\bm{\nu}}\bm{\tilde{y}} - |\tp{\bm{\nu}} \bm{\xi}| \\
 &= \tp{\bm{\nu}}\bm{\tilde{y}} - |\tp{(\diag(\bm{\ell}) \bm{\nu})} (\diag(\bm{\ell})^{-1} \bm{\xi})| \notag \\
 &\geq \tp{\bm{\nu}}\bm{\tilde{y}} - \|\diag(\bm{\ell}) \bm{\nu} \|_1 \| \diag(\bm{\ell})^{-1} \bm{\xi} \|_\infty \notag \\
 & \geq \tp{\bm{\nu}}\bm{\tilde{y}} - \|\bm{\nu} \|^{\bm{\ell}}_1, \notag
\end{align}
where $\|\bm{\nu} \|^{\bm{\ell}}_1 \equiv \|\diag(\bm{\ell}) \bm{\nu} \|_1$ denotes 
the weighted 1\nobreakdash-norm,
and we are considering the $n$-tuple $\bm{\nu}$ to be a column vector.
Hence,
\begin{equation}
\inf_{f \in D}  (\bm{\nu} \cdot \mathcal{K})[f] \geq \tp{\bm{\nu}}\bm{\tilde{y}} - \|\bm{\nu} \|^{\bm{\ell}}_1. \label{eq:lbDataTerm}
\end{equation}
When the forward functionals $\{K_j\}$ are linearly independent, 
an argument similar to that of \citet[Section 5]{Stark1992} shows that the lower bound in 
Equation~\eqref{eq:lbDataTerm} is sharp.

To summarize, we have established the inequality
\begin{equation}
  \inf_{f \in C \cap D} H_k f \geq \sup_{\bm{\nu} \in \mathbb{R}^n : \bm{\nu} \cdot \mathcal{K} \in C^*} \left\{ \tp{\bm{\nu}}\bm{\tilde{y}} - \|\bm{\nu} \|^{\bm{\ell}}_1 + \inf_{f \in C} (H_k - \bm{\nu} \cdot \mathcal{K})[f] \right\}, \label{eq:keyInequality}
\end{equation}
which holds with equality under regularity conditions. If the regularity conditions are not satisfied, the solution of the dual problem still gives a valid, conservative bound.

We now turn to the last term in Equation~\eqref{eq:keyInequality}. 
For the shape constraints we consider, $C$ is a convex cone, i.e.,
$C$ is convex and for all $\gamma \geq 0$, if $f \in C$ then 
$\gamma f \in C$. 
It then follows from \citet[Section~6.2]{Stark1992} that
\begin{align}
C^* &= \{f^* \in F^*: \inf_{f \in C} (H_k - f^*)[f] = 0 \} \\ 
&= \{f^* \in F^*: (H_k - f^*)[f] \geq 0, \;\forall f \in C \}, \notag
\end{align}
in which case the dual problem simplifies to
\begin{equation}
  \sup_{\bm{\nu} \in \mathbb{R}^n} \left\{ \tp{\bm{\nu}}\bm{\tilde{y}} - \|\bm{\nu} \|^{\bm{\ell}}_1 \right\} \quad \text{subject to} \quad (H_k - \bm{\nu} \cdot \mathcal{K})[f] \geq 0, \; \forall f \in C.
\end{equation}
The lower bound is hence given by a semi-infinite program 
with an $n$\nobreakdash-dimen\-sional free variable and an infinite set of inequality constraints.

\subsection{Constraints of the dual program} \label{sec:constraints}

We consider the following constraints $C$ on the shape of the true intensity $f$:
\begin{itemize}
 \item[(P)] $f$ is positive;
 \item[(D)] $f$ is positive and decreasing;
 \item[(C)] $f$ is positive, decreasing and convex.
\end{itemize}
The positivity constraint holds for any intensity function $f$, while monotonicity and convexity are expected to hold for steeply falling particle spectra.

We now derive the explicit form of the constraint
\begin{equation}
(H_k - \bm{\nu} \cdot \mathcal{K})[f] \geq 0,\;\; \forall f \in C,
\end{equation}
for each of these sets. 
First, re-write the left-hand side:
{\allowdisplaybreaks
\begin{align}
 \quad (H_k - \bm{\nu} \cdot \mathcal{K})[f] &= H_k f - \sum_{j=1}^n \nu_j K_j f
 = \ownint{T_k}{}{f(t)}{t} - \sum_{j=1}^n \nu_j \ownint{T}{}{k_j(t)f(t)}{t} \notag \\
 &= \ownint{T}{}{\underbrace{\bigg( \mathbf{1}_{T_k}(t) - \sum_{j=1}^n \nu_j k_j(t) \bigg)}_{\equiv h_k(t)} f(t)}{t} = \ownint{T}{}{h_k(t)f(t)}{t}. \label{eq:constraintLHS}
\end{align}
}Hence the dual constraint becomes 
$\ownint{T}{}{h_k(t)f(t)}{t} \geq 0$ 
for all $f \in C$, where 
$h_k(t) \equiv \mathbf{1}_{T_k}(t) - \sum_{j=1}^n \nu_j k_j(t)$ and 
$\mathbf{1}_{T_k}$ is the indicator function of~$T_k$.

For simplicity, assume that the space $F$ consists of twice continuously differentiable functions 
on $T$ (this assumption can be relaxed, at least for the constraints (P) and (D)). 
We furthermore assume that the kernel $k$ defined in 
Equation~\eqref{eq:kernelDef} is continuous on $S \times T$, so $\{k_j\}$ are continuous on 
$T$ and $\{h_k\}$ are right-continuous on~$T$. 
Denote the endpoints of the interval $T$ by $T_\mathrm{min}$ and $T_\mathrm{max}$. 
For each shape constraint, the dual constraint can then be equivalently written as a constraint on $h_k$:
\begin{itemize}
 \item[(P)] $h_k(t) \geq 0,\; \forall t \in T$; \smallskip
 \item[(D)] $\ownint{T_\mathrm{min}}{t}{h_k(t')}{t'} \geq 0,\; \forall t \in T$; \smallskip
 \item[(C)] $\ownint{T_\mathrm{min}}{t}{\ownint{T_\mathrm{min}}{t'}{h_k(t'')}{t''}}{t'} \geq 0,\; \forall t \in T,$ and $\ownint{T}{}{h_k(t)}{t} \geq 0$.
\end{itemize}
The result for the positivity constraint (P) follows directly, while the results for (D) and (C) 
follow from integration by parts.
The derivations are given in Appendix~\ref{app:derivationOfConstraints}. 

Define 
\begin{equation}
k_j^*(t) \equiv \ownint{T_\mathrm{min}}{t}{k_j(t')}{t'} \quad \text{and} \quad
k_j^{**}(t) \equiv \ownint{T_\mathrm{min}}{t}{\ownint{T_{\mathrm{min}}}{t'}{k_j(t'')}{t''}}{t'}.
\end{equation}
Let $\bm{\nu} \cdot \bm{k}(t) \equiv \sum_{j=1}^n \nu_j k_j(t)$ and define $\bm{\nu} \cdot \bm{k}^*(t)$ and $\bm{\nu} \cdot \bm{k}^{**}(t)$ analogously.
Substituting the definition of $h_k$ from Equation~\eqref{eq:constraintLHS} then
gives the following constraints on $\bm{\nu}$ in the optimization problem 
$\sup_{\bm{\nu} \in \mathbb{R}^n} \left\{ \tp{\bm{\nu}}\bm{\tilde{y}} - \|\bm{\nu} \|^{\bm{\ell}}_1 \right\}$:
\begin{itemize}
 \item[(P)] $\bm{\nu} \cdot \bm{k}(t) \leq L^\mathrm{P}_k(t), \; \forall t \in T$;\smallskip
 \item[(D)] $\bm{\nu} \cdot \bm{k}^*(t) \leq L^\mathrm{D}_k(t), \; \forall t \in T$;\smallskip
 \item[(C)] $\bm{\nu} \cdot \bm{k}^{**}(t) \leq L^\mathrm{C}_k(t), \; \forall t \in T$, and $\bm{\nu} \cdot \bm{k}^*(T_\mathrm{max}) \leq T_{k,\mathrm{max}} - T_{k,\mathrm{min}}$,
\end{itemize}
where the bounding functions on the right-hand side are
{\allowdisplaybreaks
\begin{align}
L^\mathrm{P}_k(t) &= \mathbf{1}_{T_k}(t), \notag \\
L^\mathrm{D}_k(t)  &= \begin{cases} 0, \; t < T_{k,\mathrm{min}}, \\ t - T_{k,\mathrm{min}}, \; T_{k,\mathrm{min}} \leq t < T_{k,\mathrm{max}}, \\ T_{k,\mathrm{max}} - T_{k,\mathrm{min}}, \; t \geq T_{k,\mathrm{max}}, \end{cases} \notag \\
L^\mathrm{C}_k(t) &= \begin{cases} 0, \; t < T_{k,\mathrm{min}}, \\ \frac{1}{2}(t - T_{k,\mathrm{min}})^2, \; T_{k,\mathrm{min}} \leq t < T_{k,\mathrm{max}}, \\ \frac{1}{2} (T_{k,\mathrm{max}} - T_{k,\mathrm{min}})^2 + (T_{k,\mathrm{max}} - T_{k,\mathrm{min}})(t - T_{k,\mathrm{max}}), \; t \geq T_{k,\mathrm{max}}. \end{cases} \notag
\end{align}}For positivity, the bounding function is piecewise constant; 
for monotonicity, it consists of two constant parts connected by a linear part; 
and for convexity, it has a constant and a linear part connected by a quadratic part.

These are explicit expressions for the constraints of the semi-infinite 
dual program corresponding to the lower bound $\underline{\lambda}_k = \inf_{f \in C \cap D} H_k f$. 
Similar reasoning shows that the upper bound
$\overline{\lambda}_k = \sup_{f \in C \cap D} H_k f= - \inf_{f \in C \cap D} -H_k f$ 
is bounded from above by  
$\inf_{\bm{\nu} \in \mathbb{R}^n} -\left\{ \tp{\bm{\nu}}\bm{\tilde{y}} - \|\bm{\nu} \|^{\bm{\ell}}_1 \right\}$ subject to the constraints:
\begin{itemize}
 \item[(P)] $\bm{\nu} \cdot \bm{k}(t) \leq -L^\mathrm{P}_k(t), \; \forall t \in T$;\smallskip
 \item[(D)] $\bm{\nu} \cdot \bm{k}^*(t) \leq -L^\mathrm{D}_k(t), \; \forall t \in T$;\smallskip
 \item[(C)] $\bm{\nu} \cdot \bm{k}^{**}(t) \leq -L^\mathrm{C}_k(t), \; \forall t \in T$, and $\bm{\nu} \cdot \bm{k}^*(T_\mathrm{max}) \leq T_{k,\mathrm{min}} - T_{k,\mathrm{max}}$.
\end{itemize}

\subsection{Discretizing the dual constraints} \label{sec:discretization}

To compute the dual bounds requires discretizing the infinite set of constraints. 
To do so, introduce a grid $\{t_i\}_{i=1}^{m+1}$ on $T$ consisting of $m+1$ grid points, with 
$m \gg p$. 
Assume that $t_1 = T_\mathrm{min}$, $t_{m+1} = T_\mathrm{max}$, 
and that there is a grid point at each boundary between the true bins $\{T_k\}_{k=1}^p$.

Imagine imposing the constraints just
at the grid points $\{t_i\}_{i=1}^{m+1}$. 
For example, the discretized version of the constraint (P) would be
\begin{equation}
 \bm{\nu} \cdot \bm{k}(t_i) \leq \pm L^\mathrm{P}_k(t_i), \quad i=1,\ldots,m+1. \label{eq:gridConstraint}
\end{equation}
This is not conservative:
the feasible set for Equation~\eqref{eq:gridConstraint} 
is \emph{larger} than that of the original constraint, so the resulting confidence interval could be
too short.

To guarantee that the actual confidence level is at least $1-\alpha$, we need to 
discretize the constraints in such a way that the 
discretized feasible set is a subset of the original feasible set for the infinite
set of constraints.
This requires taking into account the behavior of the constraints between the grid points.
We use the grid $\{t_i\}_{i=1}^{m+1}$ to find a convenient upper bound for the left-hand side of the constraint relations and then ensure that this upper bound is everywhere below the 
right-hand side functions 
$\pm L^\mathrm{P}_k$, $\pm L^\mathrm{D}_k$ or $\pm L^\mathrm{C}_k$.

\begin{figure}[t]
	\centering
	\includegraphics[trim = 0cm 0cm 0cm 0cm, clip=true, width=12.5cm]{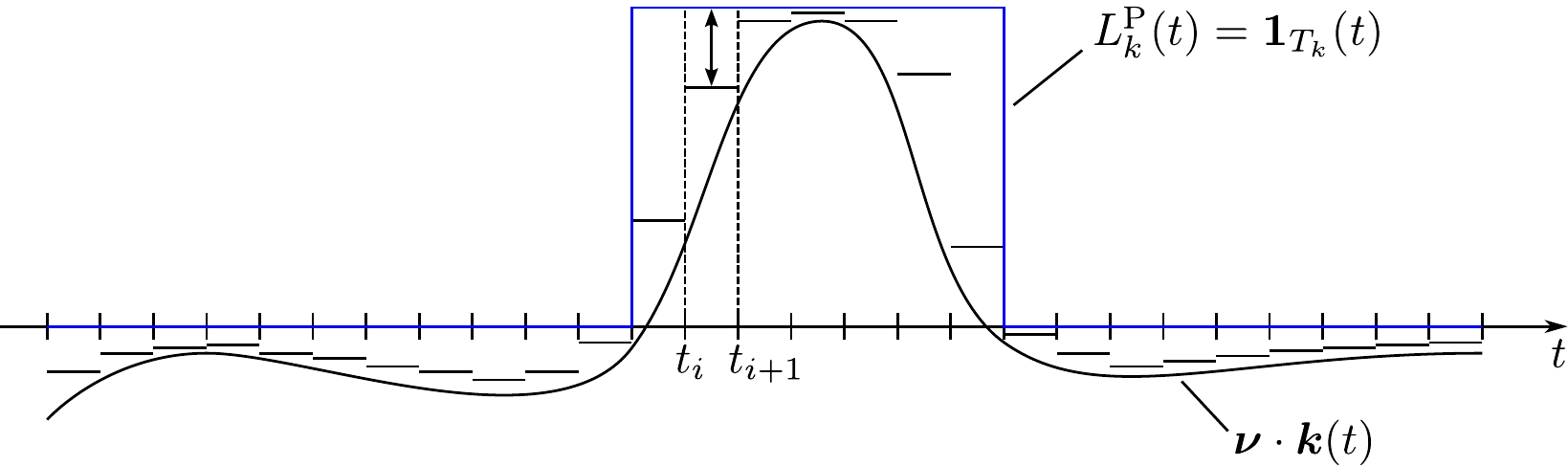}
	\caption[]{Conservative discretization of the dual constraint $\bm{\nu} \cdot \bm{k}(t) \leq L^\mathrm{P}_k(t),\, \forall t \in T$, using a constant upper bound on each interval $[t_i,t_{i+1})$.}
	\label{fig:discretization}
\end{figure}

Consider the constraint (P) for the lower bound.
For each $j=1,\ldots,n$, write $\nu_j = \nu_j^+ - \nu_j^-$ with $\nu_j^+, \nu_j^- \geq 0$,
and define the column vectors $\bm{\nu}^+$ and $\bm{\nu}^-$ in the obvious way,
so that $\bm{\nu} = \bm{\nu}^+ - \bm{\nu}^-$.
For any $t \in [t_i,t_{i+1})$,
\begin{align}
 \bm{\nu} \cdot \bm{k}(t) &= \bm{\nu}^+ \cdot \bm{k}(t) - \bm{\nu}^- \cdot \bm{k}(t) \label{eq:discreteBoundP} \\
 &\leq \sum_{j=1}^n \nu_j^+ \underbrace{\sup_{\xi \in [t_i,t_{i+1})} k_j(\xi)}_{\equiv \overline{\rho}_{i,j}} - \sum_{j=1}^n \nu_j^- \underbrace{\inf_{\xi \in [t_i,t_{i+1})} k_j(\xi)}_{\equiv \underline{\rho}_{i,j}} \notag \\
 &= \sum_{j=1}^n \nu_j^+ \overline{\rho}_{i,j} - \sum_{j=1}^n \nu_j^- \underline{\rho}_{i,j}. \notag
\end{align}This bounds $\bm{\nu} \cdot \bm{k}(t)$ on $[t_i,t_{i+1})$ by a quantity that is constant with respect to $t$.  
Since $L^\mathrm{P}_k(t)$ is also constant on $[t_i,t_{i+1})$, 
if we impose the constraints
\begin{equation}
   \sum_{j=1}^n \nu_j^+ \overline{\rho}_{i,j} - \sum_{j=1}^n \nu_j^- \underline{\rho}_{i,j} \leq L^\mathrm{P}_k(t_i), \quad i=1, \ldots, m,
\end{equation}
then the original dual constraint will hold: 
$\bm{\nu} \cdot \bm{k}(t)  \leq L^\mathrm{P}_k(t), \; \forall t \in \nobreak T$. 
Figure~\ref{fig:discretization} illustrates this construction.

Define
\begin{equation}
 \bm{A} \equiv \begin{bmatrix} \overline{\rho}_{1,1} & \cdots & \overline{\rho}_{1,n} & - \underline{\rho}_{1,1} & \cdots & -\underline{\rho}_{1,n} \\ \vdots & \ddots & \vdots & \vdots & \ddots & \vdots \\
 \overline{\rho}_{m,1} & \cdots & \overline{\rho}_{m,n} & - \underline{\rho}_{m,1} & \cdots & -\underline{\rho}_{m,n} \end{bmatrix} \quad \text{and} \quad \bm{\tilde{\nu}} \equiv \begin{bmatrix} \bm{\nu}^+ \\ \bm{\nu}^- \end{bmatrix},
\end{equation}
and let $\bm{b}_k^\mathrm{P} \in \mathbb{R}^m$ denote the column vector with components 
$b_{k,i}^\mathrm{P} = L^\mathrm{P}_k(t_i) = \mathbf{1}_{T_k}(t_i),\,i=1,\ldots,m$. 
Then the discretized dual constraint can be written~$\bm{A}\bm{\tilde{\nu}} \leq \bm{b}_k^\mathrm{P}$.

Since $\bm{\nu} = \bm{\nu}^+ - \bm{\nu}^- = \bm{D} \bm{\tilde{\nu}}$ with 
$\bm{D} \equiv \begin{bmatrix} \bm{I}_{n \times n} & -\bm{I}_{n \times n} \end{bmatrix}$ and
\begin{equation}
\|\bm{\nu}\|_1^{\bm{\ell}} = \sum_{j=1}^n \ell_j |\nu_j| \leq \sum_{j=1}^n \ell_j (\nu_j^+ + \nu_j^-) = \tp{\bm{\tilde{\ell}}\,}\bm{\tilde{\nu}} \quad \text{with} \quad \bm{\tilde{\ell}} \equiv \begin{bmatrix} \bm{\ell} \\ \bm{\ell} \end{bmatrix},
\end{equation}
any feasible point of the linear program
\vspace{-0.07cm}
\begin{equation*}
\begin{array}{cl}
 \sup\limits_{\bm{\tilde{\nu}} \in \mathbb{R}^{2n}} & \tp{(\tp{\bm{D}}\bm{\tilde{y}} - \bm{\tilde{\ell}}\,)} \bm{\tilde{\nu}} \\ \text{subject to} & \bm{A}\bm{\tilde{\nu}} \leq \bm{b}_k^\mathrm{P}, \\ & \phantom{\bm{A}} \bm{\tilde{\nu}} \geq \bm{0},
\end{array}
\vspace{-0.07cm}
\end{equation*}
yields a conservative lower bound for the $k$th element of $\bm{\lambda}_0$ 
subject to the positivity constraint (P).
Similarly, any feasible point of the linear program
\vspace{-0.07cm}
\begin{equation*}
\begin{array}{cl}
 \inf\limits_{\bm{\tilde{\nu}} \in \mathbb{R}^{2n}} & -\tp{(\tp{\bm{D}}\bm{\tilde{y}} - \bm{\tilde{\ell}}\,)} \bm{\tilde{\nu}} \\ \text{subject to} & \bm{A}\bm{\tilde{\nu}} \leq -\bm{b}_k^\mathrm{P}, \\ & \phantom{\bm{A}} \bm{\tilde{\nu}} \geq \bm{0},
\end{array}
\vspace{-0.07cm}
\end{equation*}
is a conservative upper bound.

A similar approach allows us to discretize the monotonicity constraint~(D) and the convexity 
constraint~(C) conservatively.
Deriving these discretizations is somewhat more complicated:
we use a first-order Taylor expansion of the kernels $\{k_j\}$ for~(D), and 
a second-order Taylor expansion for~(C).
For~(D), the discretized dual is again a linear program, while for~(C) the discretized
dual involves optimizing a linear 
objective function subject to a finite set of nonlinear constraints. 
Details of the discretizations of~(D) and~(C) are in Appendix~\ref{app:discretizationDC}.

\section{Simulation study} \label{sec:simulations}

\subsection{Experiment setup} \label{sec:setup}

We demonstrate the shape-constrained strict \linebreak bounds confidence intervals using a 
simulation study that mimics
unfolding the inclusive jet transverse momentum spectrum \citep{CMS2011IncJets,CMS2013IncJets} 
in the Compact Muon Solenoid (CMS) experiment \citep{CMS2008JINST} at the LHC. 
A \emph{jet} is a collimated stream of energetic particles, 
the experimental signature of a quark or a gluon created in the proton-proton collisions at the LHC. 
The jet transverse momentum spectrum describes the average number of jets as a function of their transverse momentum $\pT$, their momentum in the direction perpendicular to the proton beam.
The transverse momentum is measured in units of electron volts~(eV).
Measuring this spectrum is an important test of the 
Standard Model of particle physics and can be used to constrain free parameters of the theory.

We simulate the data using the particle-level intensity function
\begin{equation}
 f_0(\pT) = L N_0 \left(\frac{\pT}{\mathrm{GeV}}\right)^{-\alpha} \left( 1 - \frac{2}{\sqrt{s}} \pT \right)^\beta e^{-\gamma/\pT}, \quad 0 < \pT \leq \frac{\sqrt{s}}{2}. \label{eq:incJetsTrueIntensity}
\end{equation}
Here $L > 0$ is the integrated luminosity (a measure of the amount of collisions 
produced in the accelerator, measured in inverse barns, $b^{-1}$), 
$\sqrt{s}$ is the center-of-mass energy of the 
proton-proton collisions, and $N_0$, $\alpha$, $\beta$, and $\gamma$ are positive parameters. 
This parameterization is motivated by physical considerations and was used in early inclusive 
jet analyses at the LHC~\citep{CMS2011IncJets}.

When the jets are reconstructed using calorimeter information, the smearing can be modeled as additive Gaussian noise with zero mean and variance $\sigma(\pT)^2$ satisfying
\begin{equation}
\left(\frac{\sigma(\pT)}{\pT}\right)^2 = \left(\frac{C_1}{\sqrt{\pT}}\right)^2 + \left(\frac{C_2}{\pT}\right)^2 + C_3^2,
\end{equation}
where $C_i,\,i=1,2,3,$ are fixed positive constants \citep{CMS2010IncJetsPAS}. 
Let $\pT'$ denote the smeared transverse momentum.
The smeared intensity function is the convolution
\begin{equation}
 g_0(\pT') = \ownint{T}{}{N(\pT'-\pT|0,\sigma(\pT)^2)f_0(\pT)}{\pT},\quad \pT' \in S, \label{eq:incJetsFwdModel}
\end{equation}
and the unfolding problem becomes a heteroscedastic 
deconvolution problem for Poisson point process observations, 
with the forward kernel given by $k(\pT',\pT) = N(\pT'-\pT|0,\sigma(\pT)^2)$.

At the center-of-mass energy $\sqrt{s} = 7\;\mathrm{TeV}$ and in the central part of the 
CMS detector, realistic values for the parameters of $f_0(\pT)$ 
are given by $N_0 = 10^{17}\ \mathrm{fb}/\mathrm{GeV}$, $\alpha = 5$, $\beta = 10$, and $\gamma = 10\ \mathrm{GeV}$ and for the parameters of $\sigma(\pT)$ by $C_1 = 1\ \mathrm{GeV}^{1/2}$, $C_2 = 1\ \mathrm{GeV}$, and $C_3 = 0.05$ (M.~Voutilainen, personal communication, 2012). 
We furthermore set $L = 5.1\ \mathrm{fb}^{-1}$, which corresponds to the size of the CMS $7\ \mathrm{TeV}$ dataset.

The true intensity $f_0$ obviously satisfies the positivity constraint~(P). 
For the parameter values in our simulations, $f_0$ is decreasing for $\pT \gtrsim 2.0\;\mathrm{GeV}$ 
and convex for $\pT \gtrsim 2.8\;\mathrm{GeV}$. 
In other words, for intermediate and large $\pT$ values---the main focus in inclusive jet analyses \citep{CMS2013IncJets}---the true intensity satisfies the 
monotonicity constraint~(D) and the convexity constraint~(C). 
In general, physical considerations lead one to expect the $\pT$ spectrum to
satisfy these constraints, at least for intermediate values of $\pT$.

In the simulations reported here, the true and smeared spaces are 
$T = S = [400\ \mathrm{GeV},1000\ \mathrm{GeV}]$, 
and we partition both spaces into $n = p = 30$ equal-width bins. 
The dual constraints are discretized using $m + 1 = 10p + 1$ uniformly spaced grid points. 
This corresponds to subdividing each true bin into 10 sub-bins. 
The experiments were implemented in {\sc Matlab} (version~R2014a) 
using its Optimization Toolbox \citep{Mathworks2014}.
Appendix~\ref{app:implementation} gives more implementation details.
The {\sc Matlab} scripts used to produce the results are available at \url{https://github.com/mkuusela/ShapeConstrainedUnfolding}.

\subsection{Results}

\subsubsection{Shape-constrained strict bounds confidence intervals}

\begin{figure}[!hp]

\subfigure{
\includegraphics[trim = 0cm 0cm 0cm 0cm, clip=true, width=11.5cm]{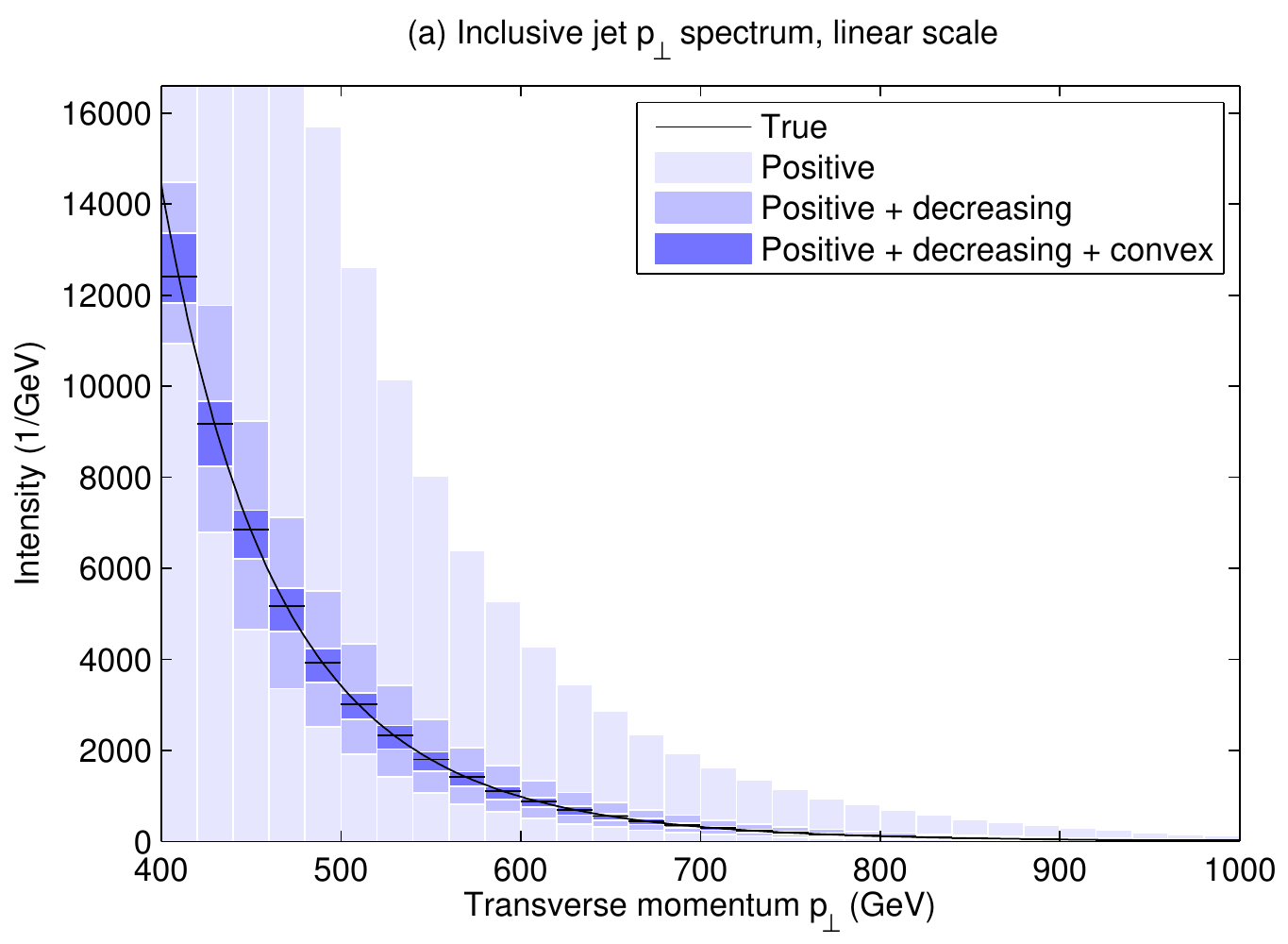}
\label{fig:strictBoundsLinear}}

\subfigure{
\includegraphics[trim = 0cm 0cm 0cm 0cm, clip=true, width=11.5cm]{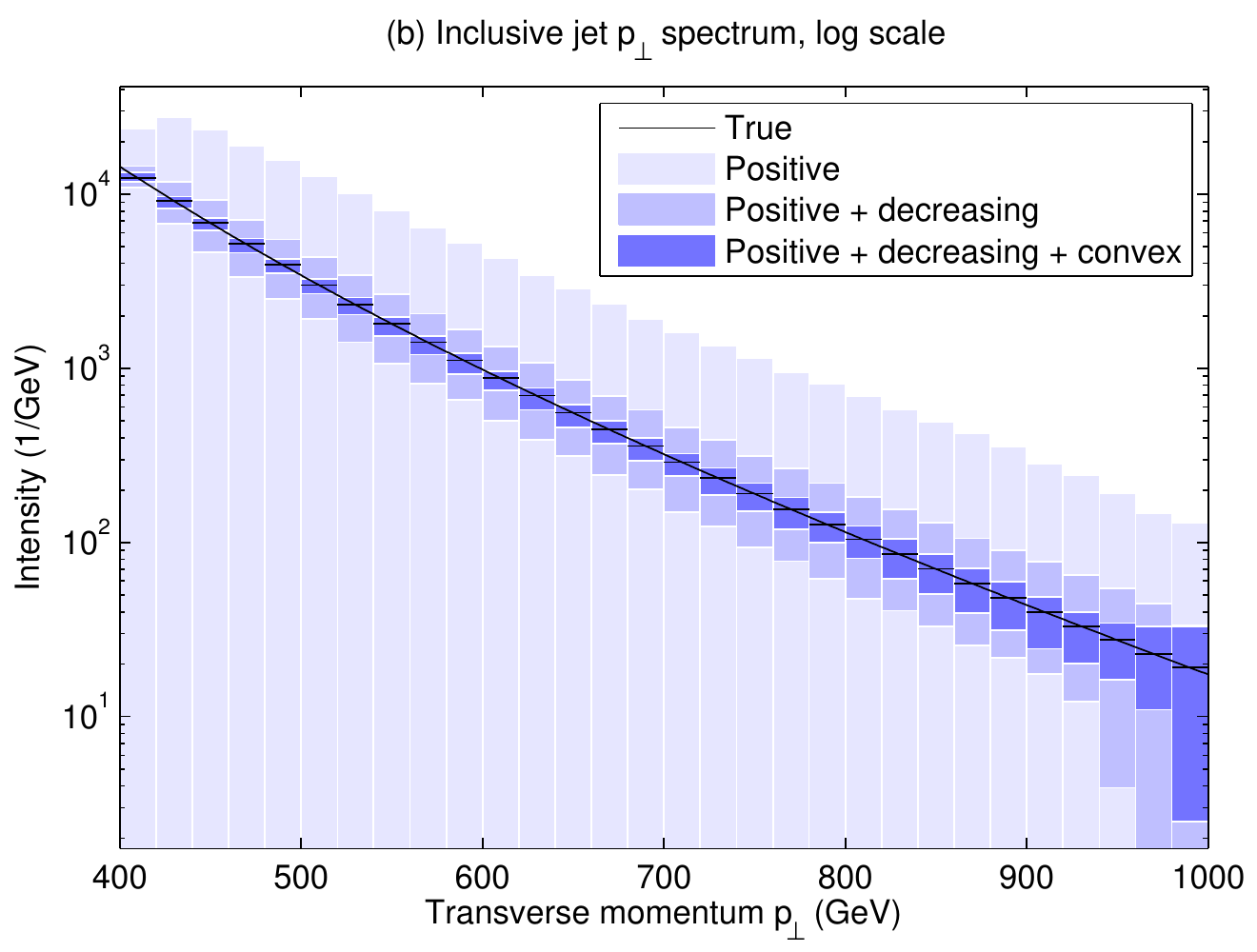}
\label{fig:strictBoundsLog}}

\caption{Shape-constrained strict bounds confidence intervals 
for the binned inclusive jet transverse momentum spectrum. 
Figure~(a) shows the results on a linear scale and Figure~(b) on a logarithmic scale. 
These intervals are guaranteed to have 95\:\% simultaneous finite-sample coverage, 
and indeed do cover the truth across the whole spectrum.}
\label{fig:strictBoundsResults}
\end{figure}

Figure~\ref{fig:strictBoundsResults} \linebreak shows the true intensity $f_0$ and the 95\:\% 
confidence intervals for $\bm{\lambda}_0$ for the different shape constraints. 
The true value of $\bm{\lambda}_0$ is shown by the horizontal lines. 
Results are plotted on both linear and logarithmic scales and the binned quantities were converted to the intensity scale by dividing them by the bin width. 
The strict bounds confidence intervals cover $\bm{\lambda}_0$ in every bin. 
The shape constraints have a marked influence on the interval lengths. 
The positivity-constrained intervals are fairly wide, 
with zero lower bound in every bin 
(they are still orders of magnitude shorter than unconstrained intervals). 
But monotonicity and convexity constraints yield much shorter intervals and 
correspondingly sharper physical inferences.

\begin{figure}[!hp]
\centering
\includegraphics[trim = 0cm 0cm 0cm 0cm, clip=true, width=12.5cm]{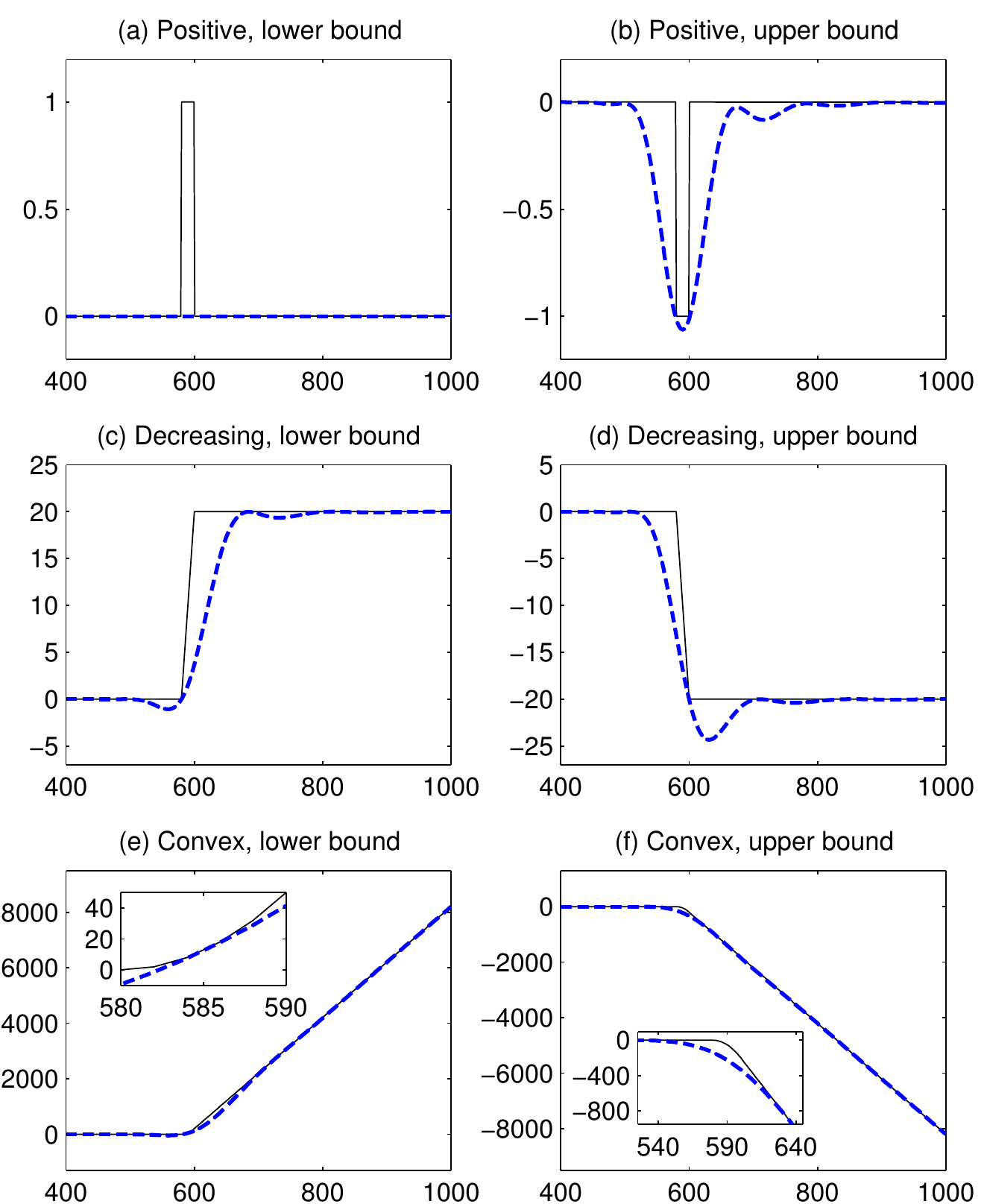}
\caption{The dual constraints $\pm L^\mathrm{P}_{10}$, $\pm L^\mathrm{D}_{10}$, and $\pm L^\mathrm{C}_{10}$ (solid lines) and the optimal solutions (dashed lines) for bounding the 10th true bin, for the three combinations of shape constraints.
Insets in Figures~(e) and (f) show the quadratic part of the constraint in greater detail.}
\label{fig:constraints}
\end{figure}

Figure~\ref{fig:constraints} shows the dual constraints $\pm L^\mathrm{P}_{10}$, $\pm L^\mathrm{D}_{10}$ and 
$\pm L^\mathrm{C}_{10}$ 
(see Section~\ref{sec:constraints}) and the corresponding optimal solutions for bounding the 10th true bin.
Despite the conservative discretization, the optimal solutions can be very close to the constraints. 
(For the positive lower bound, the optimal solution is $\bm{\tilde{\nu}} = \bm{0}$, which is consistent with the lower bound~$\underline{\lambda}_{10} = 0$.)
We also compared the lengths of the conservatively discretized intervals to those of nonconservative intervals, where the dual constraint is only imposed at the grid points $\{t_i\}_{i=1}^{m+1}$ (cf. Equation \eqref{eq:gridConstraint}). We found that the conservative intervals are only slightly wider than the nonconservative ones:
For the monotonicity and convexity constraints, the difference was less than 1\:\% in most bins. 
For the bin where the difference was the largest, the conservative intervals were 
13.2\:\%, 2.4\:\%, and 2.0\:\% longer for the positivity, monotonicity, and convexity constraints, respectively. Evidently, the conservative discretization is not excessively pessimistic.

By construction, the simultaneous coverage probability of the strict 
bounds confidence intervals in Figure~\ref{fig:strictBoundsResults} is at least 95\:\%, but in practice it can be much greater than this. 
To study the actual coverage probability, we repeated the 
study for 1,000 independent realizations of the data-generating process. 
For every realization, the intervals covered $\bm{\lambda}_0$:
in this example, the empirical simultaneous coverage probability is 100\:\% 
(the 95\:\% Clopper--Pearson confidence interval for the true
coverage probability is $[0.996,1.000]$).

To demonstrate that there are elements in the constraint set $C$ for which the intervals are less conservative, we also performed the same coverage study for a linearly decreasing intensity function
\begin{equation}
f_0(\pT) = \mathrm{const} \cdot (T_\mathrm{max} - \pT), \quad \pT \in [T_\mathrm{min}, T_\mathrm{max}], \label{eq:linDecr}
\end{equation}
and a constant intensity function
\begin{equation}
f_0(\pT) = \mathrm{const}, \quad \pT \in [T_\mathrm{min}, T_\mathrm{max}]. \label{eq:const}
\end{equation}
In both cases, $[T_\mathrm{min}, T_\mathrm{max}] = [400\ \mathrm{GeV},1000\ \mathrm{GeV}]$ and the intensity function was scaled so that the expected total number of particle-level events was the same as for the inclusive jet spectrum. 
Apart from the particle-level intensity functions, the simulation setup was the same as in the inclusive jet experiments.

To make the coverage study computationally feasible, the convexity-\linebreak constrained intervals were formed by imposing the dual constraint only on a grid; see Equation~\eqref{eq:gridConstraintCon} in Appendix~\ref{sec:shapeConstraintsDetails}. Since the intervals with the full conservative discretization are only slightly wider than these nonconservatively discretized intervals (see above), the coverage of the full intervals should not be much higher than the values reported here. The positivity- and monotonicity-constrained intervals were computed using the full conservative discretization.

\begin{table}[t]
	\caption{Empirical simultaneous coverage of the 95\:\% shape-constrained strict bounds for the inclusive jet transverse momentum spectrum of Equation~\eqref{eq:incJetsTrueIntensity}, the linearly decreasing spectrum of Equation~\eqref{eq:linDecr} and the constant spectrum of Equation~\eqref{eq:const}. The columns correspond to the different shape constraints and the rows to the different true spectra. The uncertainties in the parentheses are 95\:\% Clopper--Pearson intervals for the true coverage probability, to account for the uncertainty in
		estimating the coverage using 1,000 replications. The shape-constrained intervals are less conservative when the true spectrum is on the boundary of the constraint set.}
	\label{tab:strictBoundsCoverage}
	\begin{tabular}{llll}
		\toprule
		& \multirow{2}{*}{Positive} & Positive and & Positive, decreasing \\
		& & decreasing & and convex \\
		\midrule
		Inclusive jets & 1.000 \textit{(0.996, 1.000)} & 1.000 \textit{(0.996, 1.000)} & 1.000 \textit{(0.996, 1.000)} \\ 
		Linearly decr. & 1.000 \textit{(0.996, 1.000)} & 1.000 \textit{(0.996, 1.000)} & \textbf{0.969} \textit{(0.956, 0.979)} \\ 
		Constant & 1.000 \textit{(0.996, 1.000)} & \textbf{0.947} \textit{(0.931, 0.960)} & \textbf{0.945} \textit{(0.929, 0.958)} \\
		\bottomrule
	\end{tabular}
\end{table}

The empirical simultaneous coverage for the different intensity functions and shape constraints is given in Table~\ref{tab:strictBoundsCoverage}. 
The coverage of the monotonicity-constrained intervals is close to the nominal value when the data are generated from the constant intensity \eqref{eq:const}. 
Similarly, the coverage of the convexity-constrained intervals is close to nominal for the linearly decreasing intensity~\eqref{eq:linDecr} and the constant intensity \eqref{eq:const}. 
This suggests that the strict bounds are less conservative when the true intensity $f_0$ lies on the boundary 
of the constraint set $C$, although further studies are needed to better understand how the coverage depends on properties of $f_0$.

\subsubsection{Undercoverage of existing unfolding methods}

Currently, the two most common unfolding methods in LHC data analysis are the 
SVD variant of Tikhonov regularization \citep{Hoecker1996} 
and the D'Agostini iteration \citep{DAgostini1995}, which is an EM iteration with early stopping. 
As explained in Section~\ref{sec:intro}, these estimators are regularized by shrinking 
the solution towards a Monte Carlo prediction $\bm{\lambda}^\mathrm{MC}$ of the 
quantity of interest $\bm{\lambda}_0$. 
The methods also rely on discretizing the forward mapping using the Monte Carlo event generator. 
When the Monte Carlo prediction differs from the truth---i.e., essentially always---current 
procedures for constructing confidence 
intervals, which only account for the variance of $\bm{\hat{\lambda}}$ and not for the bias, 
can have actual coverage probabilities far lower than their nominal confidence levels. 
This section shows the severity of the problem by unfolding the 
inclusive jet $\pT$ spectrum with the SVD and D'Agostini methods.

In this simulation, the Monte Carlo event generator predicts a spectrum that is a slight perturbation of the true $f_0$:
the Monte Carlo spectrum $f^\mathrm{MC}$ is given by 
Equation~\eqref{eq:incJetsTrueIntensity}, with parameters $N_0 = 5.5 \cdot 10^{19} \ \mathrm{fb}/\mathrm{GeV}$, $\alpha = 6$, and $\beta = 12$. 
The rest of the parameters are set to the same values as before. 
This spectrum falls off slightly faster than $f_0$ in both the power-law and the energy cutoff terms. 
The value of $N_0$ was chosen so that the overall scale 
of $f^\mathrm{MC}$ is similar to $f_0$; this value has little effect on the results 
(it cancels in SVD unfolding). 
The discretized smearing matrix $\bm{K}$ and the MC prediction $\bm{\lambda}^\mathrm{MC}$ are then obtained by substituting $f^\mathrm{MC}$ into 
Equations~\eqref{eq:smearingMatrix}~and~\eqref{eq:lambda}.

We quantify the binwise uncertainty of $\bm{\hat{\lambda}}$ 
using nominal $1-\alpha$ Gaussian confidence intervals:
\begin{equation}
\left[\hat{\lambda}_k - z_{1-\alpha/2} \sqrt{\widehat{\mathrm{Var}}(\hat{\lambda}_k)},\hat{\lambda}_k + z_{1-\alpha/2} \sqrt{\widehat{\mathrm{Var}}(\hat{\lambda}_k)}\,\right],\quad k=1,\ldots,p, \label{eq:SEIntervals}
\end{equation}
where $z_{1-\alpha/2}$ is the $1-\alpha/2$ standard normal quantile and $\widehat{\mathrm{Var}}(\hat{\lambda}_k)$ is the estimated variance of $\hat{\lambda}_k$. 
This is essentially what current unfolding software implements \citep{Adye2011PHYSTAT,Schmitt2012}.
To obtain simultaneous $1-\alpha$ confidence sets for the whole histogram $\bm{\lambda}_0$, we adjust the 
levels of the binwise intervals using Bonferroni's inequality, setting the nominal level of
each interval to $1-\alpha/p$. 
In both the SVD and D'Agostini methods, the variances $\widehat{\mathrm{Var}}(\hat{\lambda}_k)$ 
are estimated using error propagation; for details, 
see Appendix~\ref{app:implementation}, which describes the two methods in more detail.

\begin{figure}[!t]

\subfigure{
\includegraphics[trim = 0.05cm 0cm 0cm 0cm, clip=true, width=6.1cm]{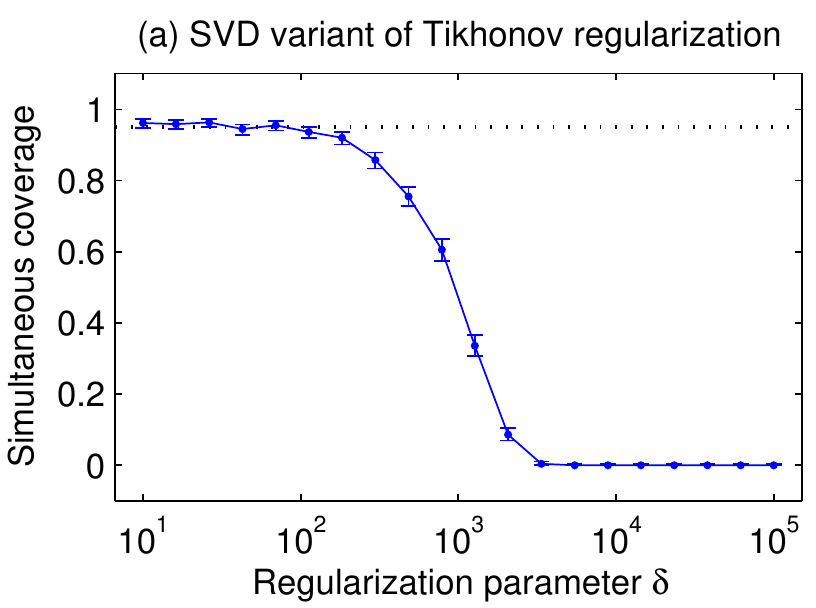}
\label{fig:SVDJointCoverage}}
\hspace{-0.2cm}
\subfigure{
\includegraphics[trim = 0.05cm 0cm 0cm 0cm, clip=true, width=6.1cm]{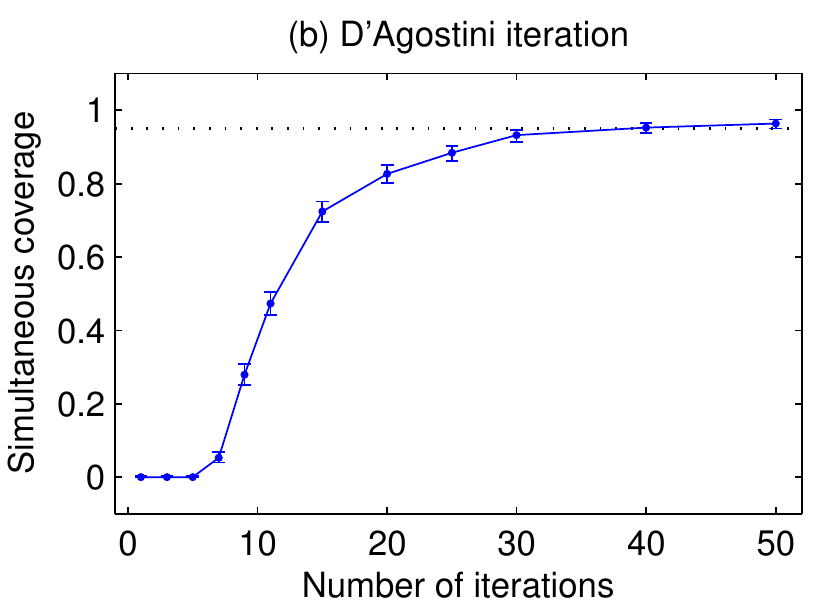}
\label{fig:DAgostiniJointCoverage}}

\caption{Simulation estimate of the simultaneous coverage of Bonferroni-adjusted nominal 95\:\% joint confidence
 intervals with (a)~the SVD variant of Tikhonov regularization and (b)~the D'Agostini iteration. 
The error bars are given by the 95\:\% 
Clopper--Pearson intervals and the nominal confidence 
level is shown by the dotted line. 
When the regularization is strong, both methods undercover substantially.}
\label{fig:SVDDAgostiniJointCoverage}
\end{figure}

The coverage properties of the intervals of 
Equation~\eqref{eq:SEIntervals} depend strongly on the regularization strength, 
which, in the case of Tikhonov regularization, is controlled by the regularization parameter $\delta$ (see Equation~\eqref{eq:Tikhonov}) and, 
in the case of the D'Agostini method, by the number of iterations. 
Figure~\ref{fig:SVDDAgostiniJointCoverage} shows the simultaneous 
coverage of (nominal) 95\:\% Bonferroni-corrected normal intervals as a function 
of the regularization strength estimated from 1,000 independent replications. 
For weak regularization, the intervals attain their nominal coverage, but, for strong regularization, 
the simultaneous coverage quickly drops to zero.

An obvious question to ask is where along these coverage curves do typical unfolding results lie?
Unfortunately, it is impossible to tell: it depends on how good the Monte Carlo predictions 
$f^\mathrm{MC}$ are and how the regularization strength is chosen. 
Currently, most analyses use nonstandard heuristics for choosing the regularization strength, without properly documenting and
justifying the criterion used. 
For instance, {\sc RooUnfold} documentation \citep{Adye2011PHYSTAT} recommends simply using four iterations for the D'Agostini method, and many LHC analyses seem to follow this convention. 
There is no principled reason to use four iterations, which in our simulations 
would result in \emph{zero} simultaneous coverage; see Figure~\ref{fig:DAgostiniJointCoverage}.

\begin{figure}[!t]

\subfigure{
\includegraphics[trim = 0cm 0cm 0cm 0cm, clip=true, width=6.1cm]{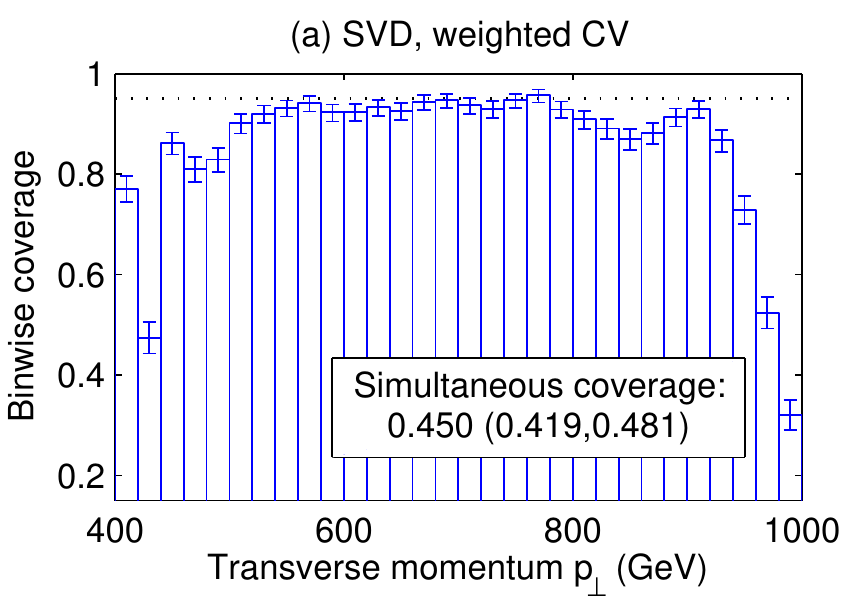}
\label{fig:SVDCVVar}}
\hspace{-0.2cm}
\subfigure{
\includegraphics[trim = 0cm 0cm 0cm 0cm, clip=true, width=6.1cm]{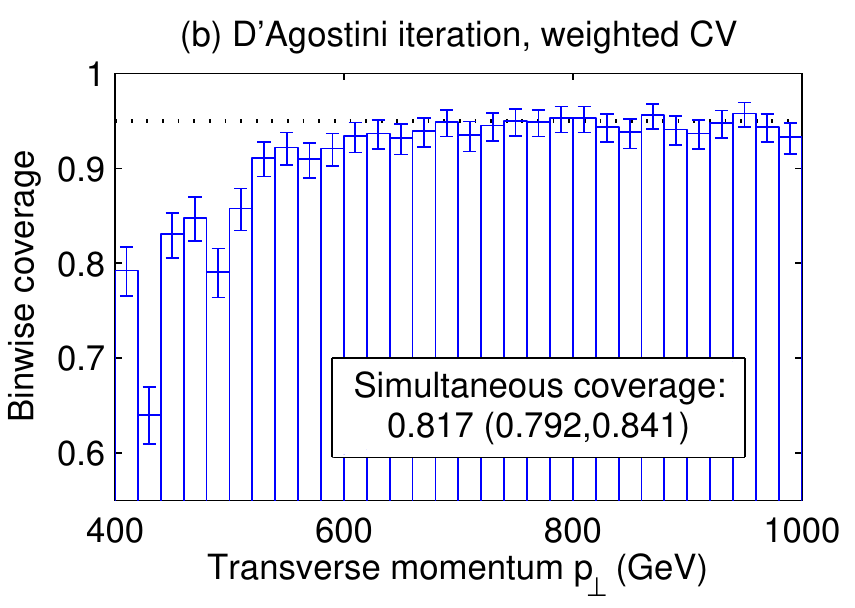}
\label{fig:DAgostiniCVVar}}

\caption{Empirical coverage of the 95\:\% normal intervals with (a)~the SVD variant of Tikhonov regularization and (b)~the D'Agostini iteration, 
when the regularization strength is chosen using weighted cross-validation. 
The simultaneous coverage is given for Bonferroni-corrected intervals.
The uncertainties are the 95\:\% Clopper--Pearson intervals for the coverage probability based on 1,000
replications. 
The nominal confidence level is shown by the dotted line. 
Both methods undercover at small values of the transverse momentum and SVD also at large transverse momenta. 
In both cases, the simultaneous coverage is below the nominal 95\:\%, dramatically so for the
SVD.}
\label{fig:SVDDAgostiniCVVar}
\end{figure}

Even with more principled, data-driven criteria for choosing the regularization strength, reasonable 
coverage performance is not guaranteed. 
Figure~\ref{fig:SVDDAgostiniCVVar} shows binwise and simultaneous empirical 
coverage of the two methods for 
1,000 independent replications when the regularization strength is selected using 
weighted cross-validation 
(see Appendix~\ref{app:implementation} for details; for D'Agostini, 3 runs where 
the cross-validation score was still decreasing after 20,000 iterations were omitted from the analysis). 
All the results are at $95\:\%$ nominal confidence. 
Both methods undercover at small $\pT$ values; the SVD approach also undercovers at 
large $\pT$ values. 
For the SVD, the smallest binwise coverage is $0.320\ (0.291,0.350)$ 
(95\:\% Clopper--Pearson interval) and, for D'Agostini, it is $0.640\ (0.609,0.670)$, 
much less than the nominal level $0.95$. 
When $f^\mathrm{MC}$ is further from $f_0$, coverage performance becomes even worse.
We also tried adding Monte Carlo noise to $\bm{\lambda}^\mathrm{MC}$, which
led to large additional reductions in the coverage.

Caution should be exercised in extrapolating these findings to 
current LHC unfolding results.
Current LHC practice treats the Monte Carlo model dependence as a systematic uncertainty. 
A typical way to take this uncertainty into account is to compute the unfolded histograms 
using two or more Monte Carlo event generators and to take the observed differences 
as an estimate of the systematic uncertainty. 
The success of this approach depends on how well the Monte Carlo models 
represent the range of plausible truths and on whether $f_0$ is in some sense 
bracketed by these models.

Clearly, the coverage of existing unfolding methods is a delicate function of the 
Monte Carlo prediction $f^\mathrm{MC}$, the number of true bins $p$, the regularization method,
the regularization strength, and the way these factors are taken into account as 
systematic uncertainties. 
It is conceivable that LHC analyses give coverage close to the nominal level,
but that would require many fortuitous accidents.
It seems impossible to give a rigorous---or even approximate---coverage guarantee for the existing methods. 
In contrast, shape-constrained strict bounds do not rely on a Monte Carlo prediction of the unknown $f_0$
and do not require the analyst to choose a regularization strength.
Moreover, the bounds are guaranteed to have nominal simultaneous coverage for any number of true 
bins~$p$, provided that $f_0$ satisfies the shape constraints, a safe assumption for 
typical steeply falling spectra.

\section{Discussion} \label{sec:discussion}

We have presented a novel approach for unfolding elementary particle spectra that
imposes physically justified shape constraints
and quantifies the uncertainty of the solution using strict bounds confidence intervals.
The resulting intervals have guaranteed simultaneous frequentist 
coverage whenever the shape constraints hold. 
This is the case for the important class of unfolding problems with 
steeply falling spectra. 
For other classes of problems, other types of shape constraints, such as unimodality, 
might still provide an attractive way to reduce uncertainty. 
A natural direction for future work would hence be to extend the approach presented 
here to $m$-modal intensities~\citep{Hengartner1995}.
Another useful extension would be to shapes that have a concave part and a convex part with
an unknown changepoint between the two.

The type of coverage considered here (simultaneous coverage for the whole histogram 
$\bm{\lambda}_0$) differs from the binwise coverage traditionally considered in~HEP. 
Binwise confidence intervals can be used to make inferential statements at a single bin, 
but interpreting the collection of such intervals as a whole is difficult. 
This issue has also been noted in the HEP literature 
\citep[Section~6.6.4]{Blobel2013}. 
In contrast, simultaneous confidence intervals, such as those in 
Figure~\ref{fig:strictBoundsResults}, by definition 
have the property that, under repeated sampling, at least 95\:\% of the envelopes 
formed by the intervals will contain the entire true histogram $\bm{\lambda}_0$. 
Hence, the confidence envelope can be directly interpreted as a whole. 

This is immediately useful for scientists who want to use unfolding results in other
analyses.
For example, a 5\:\% goodness-of-fit test of a theory prediction of $\bm{\lambda}_0$ 
can be performed by simply overlaying the prediction on the 95\:\% confidence envelope. 
(Not every spectrum within the envelope fits the data adequately, but any 
model whose predictions are not wholly
contained in the envelope can be rejected at 5\:\% significance level.)
Unfolded confidence envelopes from two independent experiments can be combined by first making an appropriate multiplicity correction and then following the strict bounds construction of Figure~\ref{fig:strictBoundsConstruction} for an identity smearing operator. 
The strict bounds construction can also be used to extract further information from the 
unfolded confidence envelopes. 
For example, one should be able to construct confidence envelopes for parton distribution functions by 
following the construction for appropriate forward operators and multiplicity-corrected unfolded input spectra. (Think of Figure~\ref{fig:strictBoundsConstruction} with the smeared space replaced by the space of unfolded inputs and the true space by the space of parton distribution functions.)
To enable this kind of re-use, it may be helpful to compute and publish unfolded confidence envelopes at 
a variety of confidence levels.

The present work assumes that the smearing kernel 
$k(s,t)$ is known perfectly, while in fact $k(s,t)$ 
is usually determined using some auxiliary measurements or simulations,
and hence is uncertain. 
If rigorous uncertainty quantification for $k(s,t)$ is available, it is possible to incorporate that uncertainty 
into the strict bounds construction \citep[Section 9.2]{Stark1992}. 
We expect this to be feasible at least when there is a physics-driven parametric model for $k(s,t)$, 
such as the calorimeter response in Equation~\eqref{eq:incJetsFwdModel}. 
Rigorous treatment of the nonparametric case, where estimating $k(s,t)$ 
essentially becomes a nonparametric quantile regression problem, appears more challenging.
Of course, these considerations also affect existing unfolding methods, 
which treat uncertainty of the smearing kernel using various heuristics.

The user is free to choose the number of smeared bins $n$ and the number of unfolded bins $p$. 
As is common in nonparametrics, it is challenging to give specific guidelines about 
how to choose them. 
However, the proposed approach guarantees conservative coverage for \emph{any} choice of $n$ and $p$, 
while the coverage of alternative nonparametric confidence intervals generally depends heavily on the choice of tuning parameters, such as a smoothing parameter or a bandwidth, as illustrated above for the SVD and D'Agostini 
methods.
Our simulations set $n = p$ following the standard practice in HEP, hence enabling a realistic comparison with existing unfolding methods. One might instead choose $n$ to be as large as available computational resources allow (particle physics data are typically not intrinsically binned). One could even eliminate $n$ altogether by constructing an unbinned confidence envelope around the smeared empirical cumulative distribution function using the Dvoretzky--Kiefer--Wolfowitz inequality as in \citet{Hengartner1992}, but this would not be computationally feasible for the large data sets typical in HEP unfolding analyses. 
In contrast, the choice of $p$ should be guided by physics judgment about the resolution at which one wishes to probe the unfolded spectrum. 
In general, the coarser the resolution (i.e., small $p$), the shorter the confidence intervals will be. 
The proposed approach is not limited to uniform bin sizes; analogous considerations apply if the bins have unequal~widths.

The proposed method potentially has large overcoverage: the intervals may be wider than necessary
to attain their nominal simultaneous confidence level.
The most important source of slack is presumably the way the set $C \cap D$ is mapped into the finite-dimensional set 
$[\underline{\lambda}_1,\overline{\lambda}_1] \times \cdots \times [\underline{\lambda}_p,\overline{\lambda}_p]$ (see Section~\ref{sec:outline}). In effect, $C \cap D$ is first mapped through $\mathcal{H}:F \rightarrow \mathbb{R}^p, f \mapsto \tp{[H_1 f,\ldots,H_p f]}$ and the resulting set $\mathcal{H}(C \cap D)$ is then bounded by a hyperrectangle. 
Shorter confidence intervals could be obtained by tuning $\Xi$ so that the geometry of $\mathcal{H}(C \cap D)$ matches better to the geometry of hyperrectangles \citep[Section 10.2]{Stark1992}. 
However, selecting $\Xi$ optimally is an open problem. Alternatively, one could consider bounding $\mathcal{H}(C \cap D)$ using some other geometric form, such as a hyperellipsoid. More generally, the optimality properties of the construction also depend on the discretization of the smeared space \citep{Hengartner1995}, and the choice to use histogram binning might not yield optimal asymptotic convergence rates.

\section*{Acknowledgments}

We warmly thank Victor Panaretos for his support of this work. 
We are also grateful to Olaf Behnke, Bob Cousins, Tommaso Dorigo, Louis Lyons, Igor Volobouev and Mikko Voutilainen for helpful discussions on unfolding, and to Yoav Zemel, the Associate Editor and the two anonymous reviewers for insightful and detailed feedback on the manuscript. 
Parts of this work were carried out while MK was visiting the Department of Statistics at the University of California, Berkeley.

\appendix

\section{Derivation of the dual constraints for decreasing and convex intensities} \label{app:derivationOfConstraints}

This section writes the constraint $\ownint{T}{}{h_k(t)f(t)}{t} \geq 0,\,\forall f \in C$, 
in an equivalent form that does not involve $f$. 
Clearly, in the case of the positivity constraint (P), an equivalent constraint is 
$h_k(t) \geq 0,\,\forall t \in T$. 
The derivations for the monotonicity constraint (D) and the convexity constraint (C) 
employ integration by parts.

\subsection{Decreasing intensities}

We show that when $C$ is the monotonicity constraint~(D),
\begin{equation}
 \ownint{T}{}{h_k(t)f(t)}{t} \geq 0,\;\forall f \in C \quad \Leftrightarrow \quad \ownint{T_\mathrm{min}}{t}{h_k(t')}{t'} \geq 0,\; \forall t \in T. \label{eq:constraintEquivalenceD}
\end{equation}
Integration by parts yields
\begin{align}
 \ownint{T}{}{h_k(t)f(t)}{t} &= \ownint{T_\mathrm{min}}{t}{h_k(t')}{t'} \, f(t) \bigg|_{T_\mathrm{min}}^{T_\mathrm{max}} - \ownint{T}{}{\ownint{T_\mathrm{min}}{t}{h_k(t')}{t'} f'(t)}{t} \notag \\
 &= \ownint{T}{}{h_k(t)}{t} \, f(T_\mathrm{max}) - \ownint{T}{}{\ownint{T_\mathrm{min}}{t}{h_k(t')}{t'} f'(t)}{t}. \label{eq:integByPartsD}
\end{align}
From this expression, it is clear that the right-hand side of
Equation~\eqref{eq:constraintEquivalenceD} implies the left-hand side. 
To show the opposite implication, assume 
that \linebreak $\ownint{T_\mathrm{min}}{t^*}{h_k(t')}{t'} < 0$ for some $t^*$ in the interior of $T$. Then, by the continuity of the integral, $\ownint{T_\mathrm{min}}{t}{h_k(t')}{t'} < 0$ for all $t \in (t^* - \delta,t^* + \delta)$ for some $\delta > 0$. 
Consider a function $d \in C$ that is a strictly positive constant on the interval 
$[T_\mathrm{min},t^*-\delta]$ and zero on $[t^*+\delta,T_\mathrm{max}]$. 
Substituting $d$ into Equation~\eqref{eq:integByPartsD} gives
\begin{equation}
 \ownint{T}{}{h_k(t)d(t)}{t} = - \ownint{t^*-\delta}{t^*+\delta}{\ownint{T_\mathrm{min}}{t}{h_k(t')}{t'} d'(t)}{t} < 0,
\end{equation}
a contradiction. 
Hence $\ownint{T_\mathrm{min}}{t}{h_k(t')}{t'} \geq 0$ for all $t$ in the interior of $T$ and, by the continuity of the integral, for all $t \in T$.

\subsection{Convex intensities}

We next treat the convexity constraint~(C) by showing that
\begin{align}
 &\ownint{T}{}{h_k(t)f(t)}{t} \geq 0,\;\forall f \in C \label{eq:constraintEquivalenceC} \\ \Leftrightarrow \quad &\ownint{T_\mathrm{min}}{t}{\ownint{T_\mathrm{min}}{t'}{h_k(t'')}{t''}}{t'} \geq 0,\; \forall t \in T \quad \wedge \quad \ownint{T}{}{h_k(t)}{t} \geq 0. \notag
\end{align}
A second application of integration by parts in Equation~\eqref{eq:integByPartsD} gives
{\allowdisplaybreaks
\begin{align}
 \ownint{T}{}{h_k(t)f(t)}{t} = \ownint{T}{}{h_k(t)}{t} \, f(T_\mathrm{max}) &- \ownint{T_\mathrm{min}}{t}{\ownint{T_\mathrm{min}}{t'}{h_k(t'')}{t''}}{t'} \, f'(t) \bigg|_{T_\mathrm{min}}^{T_\mathrm{max}} \notag \\ &+ \ownint{T}{}{\ownint{T_\mathrm{min}}{t}{\ownint{T_\mathrm{min}}{t'}{h_k(t'')}{t''}}{t'} f''(t)}{t} \notag \\
 = \ownint{T}{}{h_k(t)}{t} \, f(T_\mathrm{max}) &- \ownint{T_\mathrm{min}}{T_\mathrm{max}}{\ownint{T_\mathrm{min}}{t}{h_k(t')}{t'}}{t} \, f'(T_\mathrm{max}) \notag \\ &+ \ownint{T}{}{\ownint{T_\mathrm{min}}{t}{\ownint{T_\mathrm{min}}{t'}{h_k(t'')}{t''}}{t'} f''(t)}{t}. 
 \label{eq:integByPartsC}
\end{align}}From this form, one can see that the right-hand side of 
Equation~\eqref{eq:constraintEquivalenceC} implies the left-hand side. 
To show the reverse implication, pick $d \in C$ such that $d(t) = a > 0$ for all $t \in T$. 
Substituting this into the left-hand side implies that $\ownint{T}{}{h_k(t)}{t} \geq 0$. 
Suppose that $\ownint{T_\mathrm{min}}{t^*}{\ownint{T_\mathrm{min}}{t'}{h_k(t'')}{t''}}{t'} < 0$ for some $t^*$ in the interior of $T$. 
By the continuity of the integral, it follows that $\ownint{T_\mathrm{min}}{t}{\ownint{T_\mathrm{min}}{t'}{h_k(t'')}{t''}}{t'} < 0$ for all $t \in (t^*-\delta,t^*+\delta)$ for some $\delta > 0$.
Consider a function $d\in C$ that is linear and strictly decreasing on the interval 
$[T_\mathrm{min},t^*-\nobreak\delta]$ and zero on $[t^*+\delta,T_\mathrm{max}]$. 
Substituting $d$ into Equation~\eqref{eq:integByPartsC} yields
\begin{equation}
 \ownint{T}{}{h_k(t)d(t)}{t} = \ownint{t^*-\delta}{t^*+\delta}{\ownint{T_\mathrm{min}}{t}{\ownint{T_\mathrm{min}}{t'}{h_k(t'')}{t''}}{t'} d''(t)}{t} < 0,
\end{equation}
a contradiction.
Hence $\ownint{T_\mathrm{min}}{t}{\ownint{T_\mathrm{min}}{t'}{h_k(t'')}{t''}}{t'} \geq 0$ for all $t$ in the interior of~$T$ and, by the continuity of the integral, for all $t \in T$.

\section{Discretized constraints for decreasing and convex intensities} \label{app:discretizationDC}

This section derives conservative discretizations of the dual constraints for the monotonicity 
constraint (D) and the convexity constraint (C). 
The strategy follows the procedure of Section~\ref{sec:discretization}, with appropriate modifications. 
Discretizing the constraints 
$\bm{\nu} \cdot \bm{k}^*(t) \leq \pm L^\mathrm{D}_k(t)$ or $\bm{\nu} \cdot \bm{k}^{**}(t) \leq \pm L^\mathrm{C}_k(t)$ by applying the bound of 
Equation~\eqref{eq:discreteBoundP} to $k_j^*$ or $k_j^{**}$ gives up too much
because Equation~\eqref{eq:discreteBoundP} bounds the left-hand side by a 
constant on each interval $[t_i,t_{i+1})$, while the functions 
$\pm L^\mathrm{D}_k$ or $\pm L^\mathrm{C}_k$ on the right-hand side can vary
within these intervals. 
A better approach for the monotonicity constraint~(D) is to use a first-order 
Taylor expansion of 
$k_j^*$, which gives a linear upper bound for the left-hand side. 
For the convexity constraint~(C), we use a second-order Taylor expansion of $k_j^{**}$,
which yields a quadratic upper bound.

\subsection{Decreasing intensities} \label{sec:discretizationMonotone}

We first treat the monotonicity constraint (D). 
For any $t \in [t_i,t_{i+1})$,
\begin{align}
 k_j^*(t) &= k_j^*(t_i) + (k_j^*)'(\xi_j)(t - t_i) \\ &= k_j^*(t_i) + k_j(\xi_j)(t - t_i), \quad \xi_j \in [t_i,t). \notag
\end{align}
This yields
{\allowdisplaybreaks
\begin{align}
 \sum_{j=1}^n \nu_j k_j^*(t) &= \sum_{j=1}^n \nu_j k_j^*(t_i) + \sum_{j=1}^n \nu_j^+ k_j(\xi_j)(t - t_i) - \sum_{j=1}^n \nu_j^- k_j(\xi_j) (t - t_i) \notag \\
 &\leq \sum_{j=1}^n \nu_j k_j^*(t_i) + \sum_{j=1}^n \nu_j^+ \sup_{\xi \in [t_i,t_{i+1})} k_j(\xi) (t - t_i) \notag \\
 & \hspace{24.5mm} - \sum_{j=1}^n \nu_j^- \inf_{\xi \in [t_i,t_{i+1})} k_j(\xi) (t - t_i) \notag \\
 &= \sum_{j=1}^n \nu_j k_j^*(t_i) + \sum_{j=1}^n \nu_j^+ \overline{\rho}_{i,j} (t - t_i) - \sum_{j=1}^n \nu_j^- \underline{\rho}_{i,j} (t - t_i). \label{eq:discreteBoundD}
\end{align}}On $[t_i,t_{i+1})$, this gives a linear upper bound for 
$\bm{\nu} \cdot \bm{k}^*(t) = \sum_{j=1}^n \nu_j k_j^*(t)$. 
Since $L^\mathrm{D}_k$ is linear on each interval $[t_i,t_{i+1})$, 
it is enough to enforce the constraint at the endpoints of the interval. 
By the continuity of $L^\mathrm{D}_k$, we require for each $i=1,\ldots,m$ that
\begin{equation}
 \begin{cases}
  \sum_{j=1}^n \nu_j k_j^*(t_i) \leq \pm L^\mathrm{D}_k(t_i), \\
  \sum_{j=1}^n \nu_j k_j^*(t_i) + \sum_{j=1}^n \nu_j^+ \overline{\rho}_{i,j} \delta_i - \sum_{j=1}^n \nu_j^- \underline{\rho}_{i,j} \delta_i \leq \pm L^\mathrm{D}_k(t_{i+1}), \label{eq:discreteConstraintD}
 \end{cases}
\end{equation}
where $\delta_i \equiv t_{i+1} - t_i$. 
In fact, since $\sum_{j=1}^n \nu_j k_j^*(t)$ is continuous, the first inequality in 
Equation~\eqref{eq:discreteConstraintD} is redundant: it suffices to enforce the second.

Let $\bm{\Delta} \equiv \diag(\{\delta_i\}_{i=1}^m)$; let $\bm{K}^*$ denote the 
matrix with elements 
$K^*_{i,j} = k_j^*(t_i),\,i=1,\ldots,m,\,j=1,\ldots,n$; and let $\bm{b}^\mathrm{D}_k$ denote
the vector with elements $b^\mathrm{D}_{k,i} = L_k^\mathrm{D}(t_{i+1}),\,i=1,\ldots,m$. 
Then
$(\bm{K}^*\bm{D} + \bm{\Delta}\bm{A}) \bm{\tilde{\nu}} \leq \pm \bm{b}_k^\mathrm{D}$ is a conservative discretization of the constraint 
$\bm{\nu} \cdot \bm{k}^*(t) \leq \pm L^\mathrm{D}_k(t),\,\forall t \in T$,
where the matrices $\bm{A}$ and $\bm{D}$ are defined in Section~\ref{sec:discretization}.

Thus, any feasible point of the linear program
\begin{equation*}
\begin{array}{cl}
 \sup\limits_{\bm{\tilde{\nu}} \in \mathbb{R}^{2n}} & \tp{(\tp{\bm{D}}\bm{\tilde{y}} - \bm{\tilde{\ell}}\,)} \bm{\tilde{\nu}} \\ \text{subject to} & (\bm{K}^*\bm{D} + \bm{\Delta}\bm{A}) \bm{\tilde{\nu}} \leq \bm{b}_k^\mathrm{D}, \\ & \phantom{(\bm{K}^*\bm{D} + \bm{\Delta}\bm{A})} \bm{\tilde{\nu}} \geq \bm{0},
\end{array}
\end{equation*}
yields a conservative lower bound at the $k$th true bin. 
A conservative upper bound is given by any feasible point of
\begin{equation*}
\begin{array}{cl}
 \inf\limits_{\bm{\tilde{\nu}} \in \mathbb{R}^{2n}} & -\tp{(\tp{\bm{D}}\bm{\tilde{y}} - \bm{\tilde{\ell}}\,)} \bm{\tilde{\nu}} \\ \text{subject to} & (\bm{K}^*\bm{D} + \bm{\Delta}\bm{A}) \bm{\tilde{\nu}} \leq -\bm{b}_k^\mathrm{D}, \\ & \phantom{(\bm{K}^*\bm{D} + \bm{\Delta}\bm{A})} \bm{\tilde{\nu}} \geq \bm{0}.
\end{array}
\end{equation*}

\subsection{Convex intensities} \label{sec:discretizationConvex}
We now treat the convexity constraint (C). 
For any $t \in [t_i,t_{i+1})$,
\begin{align}
 k_j^{**}(t) &= k_j^{**}(t_i) + (k_j^{**})'(t_i)(t-t_i) + \frac{1}{2} (k_j^{**})''(\xi_j)(t-t_i)^2 \\
 &= k_j^{**}(t_i) + k_j^{*}(t_i)(t-t_i) + \frac{1}{2} k_j(\xi_j)(t-t_i)^2, \quad \xi_j \in [t_i,t). \notag
\end{align}
This yields
\begin{align}
 \sum_{j=1}^n \nu_j k_j^{**}(t) &= \sum_{j=1}^n \nu_j k_j^{**}(t_i) + \sum_{j=1}^n \nu_j k_j^{*}(t_i) (t-t_i) + \frac{1}{2} \sum_{j=1}^n \nu_j k_j(\xi_j) (t-t_i)^2 \notag \\
 &\leq \sum_{j=1}^n \nu_j k_j^{**}(t_i) + \sum_{j=1}^n \nu_j k_j^{*}(t_i) (t-t_i) \notag \\& \hspace{26.2mm} + \frac{1}{2} \sum_{j=1}^n \left( \nu_j^+ \overline{\rho}_{i,j} - \nu_j^- \underline{\rho}_{i,j} \right) (t-t_i)^2, \label{eq:discreteBoundC}
\end{align}
where the bound is established as in Equation~\eqref{eq:discreteBoundD}.
This bounds $\bm{\nu} \cdot \bm{k}^{**}(t) = \sum_{j=1}^n \nu_j k_j^{**}(t)$ on $[t_i,t_{i+1})$ from above by a parabola.
If we ensure that this parabola lies below $\pm L^\mathrm{C}_k$ for every $t \in [t_i,t_{i+1})$,
the resulting feasible set is a subset of the original set described by infinitely many constraints.
We need to require
\begin{align}
 \pm L^\mathrm{C}_k(t) &- \sum_{j=1}^n \nu_j k_j^{**}(t_i) - \sum_{j=1}^n \nu_j k_j^{*}(t_i) (t-t_i) \\ & - \frac{1}{2} \sum_{j=1}^n \left( \nu_j^+ \overline{\rho}_{i,j} - \nu_j^- \underline{\rho}_{i,j} \right) (t-t_i)^2 \geq 0, \quad \forall t \in [t_i,t_{i+1}). \notag
\end{align}
Since $L^\mathrm{C}_k$ is either linear or quadratic on $[t_i,t_{i+1})$, 
the left-hand side is a parabola and we need to ensure that this parabola is positive on the interval~$[t_i,t_{i+1})$. 

Let $a_{i,k} t^2 + b_{i,k} t + c_{i,k}$ be the parabola corresponding to the left-hand side 
and let $t^*_{i,k} \equiv - b_{i,k}/(2a_{i,k})$ be the $t$-coordinate of its vertex. 
Then $a_{i,k} t^2 + b_{i,k} t + c_{i,k} \geq 0,\,\forall t \in [t_i,t_{i+1}),$ is equivalent to requiring that
\begin{equation}
 \begin{cases}
  a_{i,k} t_i^2 + b_{i,k} t_i + c_{i,k} \geq 0, \\
  a_{i,k} t_{i+1}^2 + b_{i,k} t_{i+1} + c_{i,k} \geq 0, \\
  \mathbf{1}_{(t_i,t_{i+1})}(t^*_{i,k})(a_{i,k} (t^*_{i,k})^2 + b_{i,k} t^*_{i,k} + c_{i,k}) \geq 0.
 \end{cases} \label{eq:TaylorConstraintsC}
\end{equation}
The first two conditions guarantee that the endpoints of the parabola lie above the $t$-axis, while the last condition ensures that the vertex is above the $t$-axis if it is in the interval $(t_i,t_{i+1})$. 
As before, by the continuity of $\sum_{j=1}^n \nu_j k_j^{**}(t)$ and $L_k^\mathrm{C}(t)$, the first condition is redundant and can be dropped.

Here $t^*_{i,k}$ depends nonlinearly on $\bm{\tilde{\nu}}$, so the 
conservatively discretized optimization problem is not a linear program. 
Nevertheless, a conservative lower bound at the $k$th true bin is
\begin{equation*}
\begin{array}{cl}
 \sup\limits_{\bm{\tilde{\nu}} \in \mathbb{R}^{2n}} & \tp{(\tp{\bm{D}}\bm{\tilde{y}} - \bm{\tilde{\ell}}\,)} \bm{\tilde{\nu}} \\ \text{subject to} & a_{i,k} t_{i+1}^2 + b_{i,k} t_{i+1} + c_{i,k} \geq 0, \; i = 1,\ldots,m, \\ & \mathbf{1}_{(t_i,t_{i+1})}(t^*_{i,k})(a_{i,k} (t^*_{i,k})^2 + b_{i,k} t^*_{i,k} + c_{i,k}) \geq 0, \; i = 1,\ldots,m, \\ & -\sum\limits_{j=1}^n {(\bm{D}\bm{\tilde{\nu}})}_{\!j} \, k_j^*(T_\mathrm{max}) \geq T_{k,\mathrm{min}} - T_{k,\mathrm{max}}, \\ & \bm{\tilde{\nu}} \geq \bm{0},
\end{array}
\end{equation*}
where $\bm{\tilde{\nu}} = \begin{bmatrix} \bm{\nu}^+ \\ \bm{\nu}^- \end{bmatrix}$, $\bm{D} = \begin{bmatrix} \bm{I}_{n \times n} & -\bm{I}_{n \times n} \end{bmatrix}$, $t^*_{i,k} = - \frac{b_{i,k}}{2a_{i,k}}$ and the coefficients $a_{i,k}$, $b_{i,k}$ and $c_{i,k}$,
which depend on $\bm{\tilde{\nu}}$, are given by
\begin{align}
 a_{i,k} &= \begin{cases} -A_i, \; t_i < T_{k,\mathrm{min}}, \\ -A_i + \frac{1}{2}, \; T_{k,\mathrm{min}} \leq t_i < T_{k,\mathrm{max}}, \\ -A_i, \; t_i \geq T_{k,\mathrm{max}}, \end{cases} \\
 b_{i,k} &= \begin{cases} 2A_i t_i - B_i, \; t_i < T_{k,\mathrm{min}}, \\ 2 A_i t_i - B_i - T_{k,\mathrm{min}}, \; T_{k,\mathrm{min}} \leq t_i < T_{k,\mathrm{max}}, \\ 2 A_i t_i - B_i + T_{k,\mathrm{max}} - T_{k,\mathrm{min}}, \; t_i \geq T_{k,\mathrm{max}}, \end{cases} \\
 c_{i,k} &= \begin{cases} -A_i t_i^2 + B_i t_i - C_i, \; t_i < T_{k,\mathrm{min}}, \\ -A_i t_i^2 + B_i t_i - C_i + \frac{1}{2} T_{k,\mathrm{min}}^2, \; T_{k,\mathrm{min}} \leq t_i < T_{k,\mathrm{max}}, \\ -A_i t_i^2 + B_i t_i - C_i - \frac{1}{2} T_{k,\mathrm{max}}^2 + \frac{1}{2} T_{k,\mathrm{min}}^2, \; t_i \geq T_{k,\mathrm{max}}, \end{cases}
\end{align}
where
\begin{align}
 A_i &= \frac{1}{2} \sum_{j=1}^n \left( \nu_j^+ \overline{\rho}_{i,j} - \nu_j^- \underline{\rho}_{i,j} \right), \label{eq:defAi} \\
 B_i &= \sum_{j=1}^n (\nu_j^+ - \nu_j^-) k_j^*(t_i), \label{eq:defBi} \\
 C_i &= \sum_{j=1}^n (\nu_j^+ - \nu_j^-) k_j^{**}(t_i). \label{eq:defCi}
\end{align}
Similarly, the corresponding upper bound is 
\begin{equation*}
\begin{array}{cl}
 \inf\limits_{\bm{\tilde{\nu}} \in \mathbb{R}^{2n}} & -\tp{(\tp{\bm{D}}\bm{\tilde{y}} - \bm{\tilde{\ell}}\,)} \bm{\tilde{\nu}} \\ \text{subject to} & a_{i,k} t_{i+1}^2 + b_{i,k} t_{i+1} + c_{i,k} \geq 0, \; i = 1,\ldots,m, \\ & \mathbf{1}_{(t_i,t_{i+1})}(t^*_{i,k})(a_{i,k} (t^*_{i,k})^2 + b_{i,k} t^*_{i,k} + c_{i,k}) \geq 0, \; i = 1,\ldots,m, \\ & -\sum\limits_{j=1}^n {(\bm{D}\bm{\tilde{\nu}})}_{\!j} \, k_j^*(T_\mathrm{max}) \geq T_{k,\mathrm{max}} - T_{k,\mathrm{min}}, \\ & \bm{\tilde{\nu}} \geq \bm{0},
\end{array}
\end{equation*}
where the coefficients are
\begin{align}
 a_{i,k} &= \begin{cases} -A_i, \; t_i < T_{k,\mathrm{min}}, \\ -A_i - \frac{1}{2}, \; T_{k,\mathrm{min}} \leq t_i < T_{k,\mathrm{max}}, \\ -A_i, \; t_i \geq T_{k,\mathrm{max}}, \end{cases} \\
 b_{i,k} &= \begin{cases} 2A_i t_i - B_i, \; t_i < T_{k,\mathrm{min}}, \\ 2 A_i t_i - B_i + T_{k,\mathrm{min}}, \; T_{k,\mathrm{min}} \leq t_i < T_{k,\mathrm{max}}, \\ 2 A_i t_i - B_i - T_{k,\mathrm{max}} + T_{k,\mathrm{min}}, \; t_i \geq T_{k,\mathrm{max}}, \end{cases} \\
 c_{i,k} &= \begin{cases} -A_i t_i^2 + B_i t_i - C_i, \; t_i < T_{k,\mathrm{min}}, \\ -A_i t_i^2 + B_i t_i - C_i - \frac{1}{2} T_{k,\mathrm{min}}^2, \; T_{k,\mathrm{min}} \leq t_i < T_{k,\mathrm{max}}, \\ -A_i t_i^2 + B_i t_i - C_i + \frac{1}{2} T_{k,\mathrm{max}}^2 - \frac{1}{2} T_{k,\mathrm{min}}^2, \; t_i \geq T_{k,\mathrm{max}}, \end{cases}
\end{align}
and $A_i$, $B_i$ and $C_i$ are given by Equations~\eqref{eq:defAi}--\eqref{eq:defCi}.

These optimization problems can be solved using standard nonlinear programming algorithms, 
but some care is needed when choosing the algorithm and its starting point; see Section~\ref{sec:shapeConstraintsDetails}. 
Since any feasible point of these programs yields a conservative bound, 
we need only a good feasible point and not necessarily a global optimum.

\section{Implementation details} \label{app:implementation}

This appendix provides implementation details for the unfolding methods used in this paper.
Section~\ref{sec:shapeConstraintsDetails} focuses on the shape-constrained strict bounds, 
while Sections~\ref{sec:SVDDetails} and \ref{sec:DAgostiniDetails} 
describe our implementation of the SVD and D'Agostini methods, respectively.

\subsection{Shape-constrained strict bounds} \label{sec:shapeConstraintsDetails}

The optimization problems to find the shape-constrained strict bounds involve a relatively high-di\-men\-sion\-al solution space, 
numerical values on very different scales, and fairly complicated constraints. 
As a result, some care is needed in their numerical solution, including verifying
the validity of the output of the optimization algorithms.

For positivity and monotonicity constraints, where the bounds can be 
found by linear programming, we used the interior-point linear program solver 
implemented in the \texttt{linprog} function of the {\sc Matlab} 
Optimization Toolbox R2014a \citep{Mathworks2014}. 
To find bounds under the convexity constraint,
we used the sequential quadratic programming (SQP) algorithm as implemented in the 
\texttt{fmincon} function of the {\sc Matlab} Optimization Toolbox R2014a \citep{Mathworks2014}. 
(Due to a change in our computing infrastructure, the simulations for 
Table~\ref{tab:strictBoundsCoverage} were run using {\sc Matlab} version R2014b, but we do not expect
that to affect the results.)

The optimization problems described in Sections~\ref{sec:discretization}, 
\ref{sec:discretizationMonotone}, and \ref{sec:discretizationConvex} 
tend to be numerically unstable when the solver explores large values of $\bm{\tilde{\nu}}$. 
We address this issue by imposing an upper bound on~$\bm{\tilde{\nu}}$. 
For each $j$, we replace the constraint $\tilde{\nu}_j \geq 0$ with the constraint 
$0 \leq \tilde{\nu}_j \leq U$, where $U$ is chosen to be large enough 
that the upper bound is not active at the optimal solution. 
(Notice that even if the upper bound were active, the solution of the modified problem
would still be a valid confidence bound, 
because the restricted feasible set is a subset of the original feasible set.) 
Imposing the upper bound substantially improved the stability of the numerical solvers. 
For the numerical experiments of this paper, $U$ was set to 30 for the positivity constraint, 
15 for the monotonicity constraint, and 10 for the convexity~constraint.

The solution found by the optimization algorithms can violate the constraints within a 
preset numerical tolerance.
This could make the confidence bound optimistic rather than conservative.
To ensure that this does not happen, we verify the feasibility of the solution returned by the optimization algorithm.
If it is infeasible, we iteratively scale $\bm{\nu}^+$ down and $\bm{\nu}^-$ up 
until the solution becomes feasible. 
Typically very little fine-tuning of this kind was required to obtain a feasible point.

The SQP algorithm requires a good feasible starting point. 
To find one, we first solve the linear program corresponding to the 
nonconservative discretization
\begin{equation}
 \bm{\nu} \cdot \bm{k}^{**}(t_i) \leq \pm L^\mathrm{C}_k(t_i), \quad i=1,\ldots,m+1. \label{eq:gridConstraintCon}
\end{equation}
We then scale the solution as described above to make it feasible for the 
conservative discretization; the result is then used as the initial feasible point for SQP. 
Since the SQP iteration is prohibitively CPU intensive for a large-scale coverage study, the nonconservative discretization \eqref{eq:gridConstraintCon} was also used to obtain the convexity-constrained intervals for the coverage study of Table~\ref{tab:strictBoundsCoverage}.

The implementation described here worked well for the inclusive jet spectrum of 
Section~\ref{sec:simulations}.
Occasionally, the algorithms returned a suboptimal feasible point. 
This maintains conservative coverage, but adjusting the tuning parameters of the 
optimization algorithms might help find a better feasible point.
For the lower bound, the point $\bm{\tilde{\nu}} = \bm{0}$ is always feasible 
(yielding a lower bound
of zero), while, for the upper bound, the algorithms might find no feasible point, in 
which case the upper confidence bound is~$+\infty$.

\subsection{SVD variant of Tikhonov regularization} \label{sec:SVDDetails}

The SVD unfolding technique of \citet{Hoecker1996} is a variant of 
Tikhonov \linebreak regularization: the unfolded estimator $\bm{\hat{\lambda}}$ in 
the discretized model \linebreak $\bm{y} \sim \mathrm{Poisson}(\bm{K}\bm{\lambda}_0)$ 
solves the optimization problem
\begin{equation}
\min_{\bm{\lambda} \in \mathbb{R}^p} \tp{(\bm{y}-\bm{K}\bm{\lambda})} \bm{\hat{C}}^{-1} (\bm{y}-\bm{K}\bm{\lambda}) + \delta \|\bm{\tilde{L}}\bm{\lambda}\|_2^2, \label{eq:SVDOptProb}
\end{equation}
where $\bm{\hat{C}} \equiv \mathrm{diag}(\bm{y})$ is the estimated covariance of $\bm{y}$ and $\bm{\tilde{L}} \equiv \bm{L}\,\mathrm{diag}(\bm{\lambda}^\mathrm{MC})^{-1}$, with
\begin{equation}
 \bm{L} = \begin{bmatrix} -1 & 1 &&&& \\ 1 & -2 & 1 &&& \\ & 1 & -2 & 1 && \\ && \ddots & \ddots & \ddots & \\ &&& 1 & -2 & 1 \\ &&&& 1 & -1 \end{bmatrix}.
\end{equation}
The matrix $\bm{L}$ corresponds to a discretized second derivative with reflexive 
boundary conditions \citep{Hansen2010}.
In other words, the problem is regularized by penalizing the second derivative 
of the binwise ratio of the 
unfolded histogram $\bm{\lambda}$ and its MC prediction $\bm{\lambda}^\mathrm{MC}$.

The solution to Equation~\eqref{eq:SVDOptProb} 
is the point estimator
\begin{equation}
 \bm{\hat{\lambda}} = \underbrace{(\tp{\bm{K}} \bm{\hat{C}}^{-1} \bm{K} + \delta \tp{\bm{\tilde{L}}}\bm{\tilde{L}})^{-1} \tp{\bm{K}} \bm{\hat{C}}^{-1}}_{\equiv \bm{K}^+} \bm{y} = \bm{K}^+ \bm{y}.
\end{equation}
The corresponding estimate of the smeared histogram $\bm{\mu}_0$ is $\bm{\hat{\mu}} = \bm{K} \bm{\hat{\lambda}} = \bm{K} \bm{K}^+ \bm{y} = \bm{H} \bm{y}$ with $\bm{H} \equiv \bm{K} \bm{K}^+$. Since the variance of $\bm{y}$ differs across the bins, 
we select the regularization parameter $\delta$ using weighted leave\nobreakdash-one\nobreakdash-out cross-validation 
\citep[Section 3.5.3]{Green1994} weighting each bin 
by the reciprocal of its estimated variance. That is, we choose $\delta$ to minimize
\begin{equation}
 \mathrm{CV} = \sum_{i=1}^n \frac{(y_i - \hat{\mu}_i^{-i})^2}{y_i} = \sum_{i=1}^n \frac{1}{y_i} \left(\frac{y_i - \hat{\mu}_i}{1-H_{i,i}}\right)^2, \label{eq:CVFunction}
\end{equation}
where $\hat{\mu}_i^{-i}$ is the estimate of the $i$th smeared bin obtained using $\bm{y}^{-i} = \tp{[y_1,\ldots,y_{i-1},y_{i+1},\ldots,y_n]}$.
This choice of $\delta$ aims to minimize the squared prediction error in the smeared space, as described for example in \citet{OSullivan1986}. 
\citet[Chapter~5]{Ruppert2003}, \citet[Chapter~7]{Hansen1998}, and \citet[Chapter~7]{Vogel2002}
review alternative techniques for choosing the regularization parameter.

An estimate of the covariance of $\bm{\hat{\lambda}}$ (ignoring the data-dependence 
of $\delta$ and~$\bm{\hat{C}}$)
is  $\widehat{\mathrm{Cov}}(\bm{\hat{\lambda}}) = \bm{K}^+ \mathrm{diag}(\bm{y}) \tp{(\bm{K}^+)}$. 
If the distribution of $\bm{\hat{\lambda}}$ is approximately Gaussian, its binwise 
uncertainties can be quantified using the intervals of
Equation~\eqref{eq:SEIntervals} with 
$\widehat{\mathrm{Var}}(\hat{\lambda}_j) = \widehat{\mathrm{Cov}}(\bm{\hat{\lambda}})_{j,j}$.

\subsection{D'Agostini iteration} \label{sec:DAgostiniDetails}

D'Agostini iteration \citep{DAgostini1995} uses the EM algorithm \citep{Dempster1977,Shepp1982,Vardi1985} to find a maximum 
likelihood solution in the discrete forward model 
$\bm{y} \sim \mathrm{Poisson}(\bm{K}\bm{\lambda}_0)$. 
Given a starting point $\bm{\lambda}^{(0)} > \bm{0}$, the $(t+1)$th step of the 
iteration is
\begin{equation}
 \lambda_j^{(t+1)} = \frac{\lambda_j^{(t)}}{\sum_{i=1}^n K_{i,j}} \sum_{i=1}^n \frac{K_{i,j}y_i}{\sum_{k=1}^p K_{i,k} \lambda_k^{(t)}}.
\end{equation}
The solution is regularized by stopping the algorithm before it converges to a maximum likelihood estimate. 
In the {\sc RooUnfold} implementation \citep{Adye2011PHYSTAT}, the iteration starts at
a MC prediction of the unfolded histogram, $\bm{\lambda}^{(0)} = \bm{\lambda}^\mathrm{MC}$.

We choose the number of iterations using weighted cross-val\-i\-da\-tion, minimizing Equation~\eqref{eq:CVFunction}. Due to the nonlinearity of the D'Agostini iteration, the second equality in Equation~\eqref{eq:CVFunction} no longer holds.
As a result, evaluating the cross-validation function requires computing $n$ point estimates, 
one for each left-out smeared bin, which is computationally demanding.

Using linearization, we estimate the covariance 
of $\bm{\lambda}^{(t+1)}$ by
\begin{equation}
\widehat{\mathrm{Cov}}(\bm{\lambda}^{(t+1)}) = \bm{J}^{(t+1)} \mathrm{diag}(\bm{y}) \tp{(\bm{J}^{(t+1)})},
\end{equation}
where $\bm{J}^{(t+1)}$ is the Jacobian of $\bm{\lambda}^{(t+1)}$ evaluated at $\bm{y}$. 
Let $\varepsilon_j \equiv \sum_{i=1}^n K_{i,j}$ and
\begin{equation}
M_{i,j}^{(t)} \equiv \frac{\lambda_j^{(t)}}{\varepsilon_j} \frac{K_{i,j}}{\sum_{k=1}^p K_{i,k} \lambda_k^{(t)}}.
\end{equation}
Then the elements of the Jacobian are \citep{Adye2011PHYSTAT}
\begin{equation}
 J_{j,i}^{(t+1)} = \frac{\partial \lambda_j^{(t+1)}}{\partial y_i} = M_{i,j}^{(t)} + \frac{\lambda_j^{(t+1)}}{\lambda_j^{(t)}} J_{j,i}^{(t)} - \sum_{k=1}^p \sum_{l=1}^n y_l \frac{\varepsilon_k}{\lambda_k^{(t)}} M_{l,j}^{(t)} M_{l,k}^{(t)} J_{k,i}^{(t)},
\end{equation}
with $J_{j,i}^{(0)} = 0$ for all $j,i$. 
The estimated variances for Equation~\eqref{eq:SEIntervals} are 
$\widehat{\mathrm{Var}}(\hat{\lambda}_j) = \widehat{\mathrm{Cov}}(\bm{\hat{\lambda}})_{j,j} = \widehat{\mathrm{Cov}}(\bm{\lambda}^{(N_\mathrm{iter})})_{j,j}$, where $N_\mathrm{iter}$ is the number of iterations.
The resulting intervals ignore the data-dependence of $N_\mathrm{iter}$ and treat $\bm{\hat{\lambda}}$ 
as Gaussian.


\bibliographystyle{imsart-nameyear}
\bibliography{references}

\end{document}